\documentclass{ws-ijmpd}
\usepackage[super,compress]{cite}
\usepackage{hyperref}
\usepackage{amssymb}
\usepackage{amsmath}
\usepackage{graphicx}
\usepackage{esvect}
\usepackage{multicol}
\usepackage{multirow}
\usepackage{array}
\usepackage[utf8]{inputenc}
\usepackage[square,sort,comma,numbers]{natbib}
\usepackage{floatrow}
\usepackage{natbib}
\usepackage{gensymb}
\usepackage{makecell}
\usepackage{lipsum}
\usepackage{pdflscape}
\usepackage{rotating}
\usepackage{capt-of}
\usepackage{color}

\begin{document}

\markboth{S.B et.al}
{Interpretation of CALET $e^+\, +\, e^-$ Data}

%
\catchline{}{}{}{}{}
%

\title{An Interpretation of the Cosmic Ray $e^++e^-$ Spectrum from 10~GeV to 3~TeV Measured by CALET on the ISS}

\author{Saptashwa Bhattacharyya}

\address{Graduate School of Advanced Science and Engineering, Waseda University\\
Shinjuku, Tokyo,
Japan \\
saptashwab@ruri.waseda.jp}

\author{Holger Motz} 

\address{Global Center for Science and Engineering, Waseda University\\
Shinjuku, Tokyo, Japan\\
motz@aoni.waseda.jp}

\author{Yoichi Asaoka}

\address{Waseda Research Institute for Science and Engineering, Waseda University\\
Shinjuku, Tokyo, Japan\\
yoichi.asaoka@aoni.waseda.jp}

\author{Shoji Torii}
\address{Graduate School of Advanced Science and Engineering, Waseda University\\
Waseda Research Institute for Science and Engineering, Waseda University\\
Shinjuku, Tokyo, Japan\\
shoji.torii@waseda.jp}

\maketitle

\begin{history}
\received{Day Month Year}
\revised{Day Month Year}
\end{history}

\begin{abstract}
 A combined interpretation of the CALET $e^+\, +\, e^-$ spectrum up to 3~TeV and the AMS-02 positron spectrum up to 500~GeV was performed and the results are discussed. To parametrize the background electron flux, we assume a smoothly broken power-law spectrum with an exponential cut-off for electrons and fit this parametrization to the measurements, with either a pulsar or 3-body decay of fermionic Dark Matter as the extra electron-positron pair source responsible for the positron excess. We found that depending on the parameters for the background, both Dark Matter decay and the pulsar model can explain the combined measurements. While the Dark Matter decay scenario is constrained by the Fermi-LAT $\gamma$-ray measurement, we show that 3-body decay of a 800~GeV Dark Matter can be compatible with the $\gamma$-ray flux measurement. 
We discuss the capability of CALET to discern decaying Dark Matter models from a generic pulsar source scenario, based on simulated data for five~years of data-taking.

\end{abstract}


\ccode{PACS numbers:}


\newpage
\section{Introduction}
Precise measurements from space-based detectors have led to significant progress in the understanding of the cosmic-ray (CR) fluxes in the last decade. The magnetic spectrometers PAMELA~\cite{Casolino:Pamelalaunch} and AMS-02~\cite{Aguilar:AMSfrac} provided accurate measurements of the CR electron and positron spectra up to several 100 GeV, discovering and verifying the positron excess. The CALorimetric Electron Telescope (CALET), in operation on the International Space Station (ISS) since October 2015, features excellent energy resolution $(2\%)$ and \mbox{proton rejection power $(1:10^5)$}, measuring the combined \mbox{$e^+\, +\, e^-$} spectrum up to 20 TeV~\cite{Tori:2017saw, Asaoka:calib}. Recently, the CALET collaboration has published first results of the CR \mbox{$e^+\, +\, e^-$} spectrum based on \mbox{627 days} of operation, extending the measured energy range up to 3~TeV~\cite{PhysRevLett:CALET}. 

One of the most important features of the CR electron and positron spectra is the positron excess observed by PAMELA~\cite{Adriani:PAMELAfrac} and AMS-02~\cite{Aguilar:AMSfrac, Accardo:AMSfrac} which could be explained by a nearby pulsar, or more exotic sources producing electron-positron pairs such as annihilation or decay of Dark Matter (DM)~\cite{Bregstrom:DMposi, Cholis:DMposi, DMdecayposi:China, Hooper:DMpulsarplot, Feng:2015uta, Hooper:Pulsarposi}. The precise measurement of the CR \mbox{$e^+\, +\, e^-$}  spectrum at high energies by CALET can be used to investigate the type of this extra source, since specific spectral features are expected near the highest energy of the particles originating from the extra source \cite{Bhattacharyya:selfcite}. In this context, we discuss DM decay as a possible extra source that can provide DM-only explanation of the positron excess, specifically the scenario in which DM decays into two charged Standard Model (SM) leptons and a neutrino. Alternatively, we study the generic power-law with cut-off extra source from the minimal model proposed by the AMS-02 collaboration~\cite{Aguilar:AMSfrac} as an explanation of their positron fraction measurement. This source spectrum corresponds well to that of a single young pulsar~\cite{Motz:2015cua}, and is thus associated with the pulsar interpretation of the positron excess.

Both the DM decay and pulsar scenario are tested with the combined \mbox{$e^+\, +\, e^-$} spectrum and $e^+$ spectrum measured by CALET and AMS-02 respectively. For this comparison, a parametrization of the local interstellar spectra is fitted to both the $e^+\, +\, e^-$ and $e^+$ data. It is shown that depending on the choice of parameters for modeling the background flux, both the DM decay and pulsar model can well explain the measurements. As a result, we derive the allowed ranges for the pulsar-spectrum cut-off energy and the DM mass at a confidence level (CL) of $95\%$. It should be noted that we do not include the systematic errors of the CALET measurement in this analysis in order to investigate the possible fine structures in the spectrum. Also, significant components of the systematic errors are uncertainties on the flux normalization which do not affect the spectral feature study. The systematics are expected to be reduced for the five-year observation data~\cite{PhysRevLett:CALET}, allowing for a conclusive test of these results.         

Since the decay of DM is accompanied by $\gamma$-ray emission, selected DM decay scenarios well in agreement with the $e^+\, +\, e^-$ and $e^+$ spectra are compared with the results of the Fermi-LAT $\gamma$-ray observation~\cite{Ackermann:fermigammaspectrum}. DM with a mass of 800~GeV is shown to be compatible with the Fermi-LAT $\gamma$-ray measurement. 
Furthermore, based on simulated data, it was investigated to what extent the selected well fitting models of DM decay, for which also the $\gamma$-ray emission was calculated, could be distinguished from a single pulsar extra source with five-years of CALET data. It is shown that depending on DM decay and background parameters, a good chance for separating the two cases exists.   

\begin{figure}
 \centering 
 \includegraphics[width=1.\linewidth]{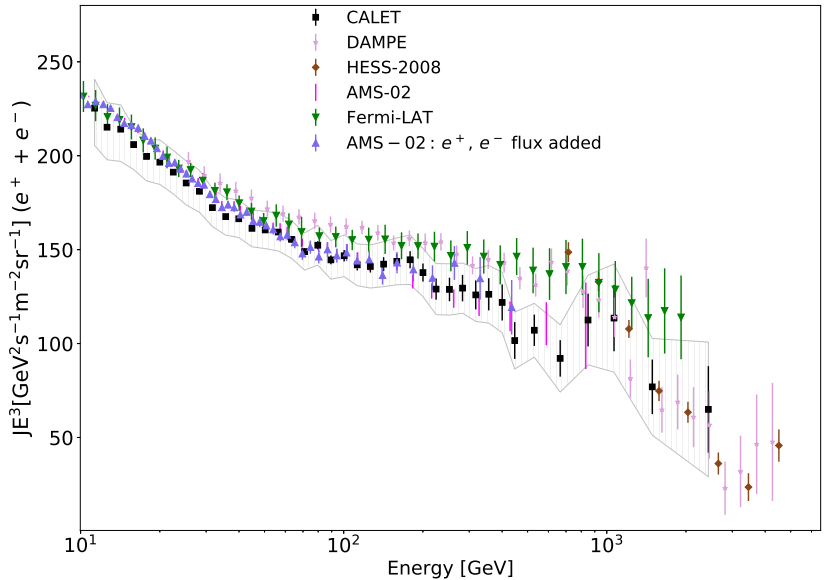}
\caption{Results of cosmic-ray $e^+\, +\, e^-$ spectrum measured by recent space-based~\cite{PhysRevLett:CALET, DAMPEnature, Abdollahi:2017nat, Aguilar:2014totalflux} and a ground based experiment~\cite{Aharonian:2008aa} are shown here. Systematic errors of CALET are shown as a grey band. \label{fig:overviewtotalflux}}
\end{figure}

While this paper deals exclusively with the interpretation of the CALET measurement, several other space based detectors and also air-Cherenkov telescopes have measured the CR $e^+\, +\, e^-$ spectrum in a similar energy range. An overview of these results of AMS-02~\cite{Aguilar:2014totalflux}, DAMPE~\cite{DAMPEnature}, Fermi-LAT~\cite{Abdollahi:2017nat}, HESS (2008)~\cite{Aharonian:2008aa} is given in figure~\ref{fig:overviewtotalflux}. The flux between a few tens of GeV and around 1~TeV reported by \mbox{Fermi-LAT} and DAMPE is systematically higher than that of AMS-02 and CALET, which are compatible within errors. While above around 1~TeV the experimental results show better agreement, 
the disputed several hundred~GeV energy range is most important for interpretation of the positron excess. A comparative analysis with DAMPE data could not be done, since using the same method with DAMPE and AMS-02 data combined, no parameter space remains for the 3-particle DM decay scenario as an explanation of the positron excess.
While the discrepancy between CALET and DAMPE results is not resolvable without taking possible systematic shifts into account, CALET data is given preference, since our combined analysis requires consistency between the $e^+$ and $e^+\,+\, e^-$  spectra, and the DAMPE result is not well consistent with the AMS-02 \mbox{$e^+\, +\, e^-$} measurement within the energy range of our analysis. Though the AMS-02 \mbox{$e^+\, +\, e^-$}  flux and $e^+$~flux are derived from independent analyses, the sum of $e^+$ and $e^-$ flux is in agreement with the \mbox{$e^+\, +\, e^-$} flux and CALET data, as shown in figure~\ref{fig:overviewtotalflux}, indicating that no systematic shifts have to be taken into account in the combined analysis. 

This paper is organized as follows: First, in section~\ref{sec:param}, the parametrization of the local CR $e^-$ and $e^+$ spectra is described. The fitting results with DM decay and pulsar as extra sources are described in section~\ref{sec:DMfit} and section~\ref{sec:pulfit} respectively. Finally, in section~\ref{sec:5yrCALET} we discuss the discerning capability of CALET between these two potential extra sources of positrons with five years of CALET data.

\section{Parametrization of the Cosmic-Ray $e^-$ and $e^+$ Spectra}\label{sec:param}

It is necessary to parametrize the local interstellar $e^+$ and $e^-$ spectra with analytical formulae in order to reflect their variability caused by the unknown free parameters of injection and propagation. This flexible modeling of the background is required for an unbiased study of the possible extra source contributions to the observed spectrum. 

A parametrization suitable to describe the AMS-02 $e^+\, +\, e^-$ and $e^+$ spectra was introduced in our previous works \cite{Motz:2015cua,Bhattacharyya:selfcite}. However, it was found that a single power law function for the primary electron flux does not well describe the CALET $e^+\, +\, e^-$ spectrum (including only statistical error) in combination with the AMS-02 $e^+$ spectrum, with $\chi ^2$ of the fit clearly above $95\%$ CL.

Since both the electron spectrum and the positron spectrum measured by \mbox{AMS-02} show a spectral hardening (spectral index changing from $-2.97\pm 0.03$ to $-2.75\pm 0.05$ for positrons and spectral index changing from $-3.28\pm 0.03$ to $-3.15\pm 0.04$ for electrons)~\cite{Aguilar:posielec}, we include a spectral break in the electron flux parametrization to match the measurements over a wide energy range.

The primary electron spectrum also features an exponential cut-off representing the radiative energy loss processes during propagation in the interstellar medium (ISM), since a discrete distribution of supernova remnants (SNRs) contributing to the primary electron spectrum is assumed, causing primary electrons to have propagated at least the distance from the closest SNR. This cut-off term is not present for the secondaries, represented by another \mbox{power-law} term with index reduced by $\delta$, the power-law index of the diffusion coefficient, since they are products of nuclei interaction with the ISM, and energy loss processes are regarded as negligible for nucleons compared to electrons.  
To this background, either the DM decay or the pulsar source providing extra electron-positron pairs is added.

In total, the electron flux can be expressed as 
\begin{equation}
 \label{eq:bplelec}
\phi_{e^-} = C_{e^-}E^{\gamma _{e^-} - \Delta\gamma _{e^-}}\left(1+\left(\frac{E}{E_g}\right)^{\frac{\Delta\gamma _{e^{-}}}{s}}\right)^{s}\, \left(\frac{C_s}{C_{e^-}}\, E^{-\gamma _{e^-}+\gamma _s}\, + e^{-E/E_d}\right) +\, \phi _{\text{extra}}\, \, \, ,
\end{equation}
where $C_{e^-},\, \frac{C_s}{C_ {e^-}}$ are the absolute normalization of the primary $e^-$ flux and the ratio of absolute normalization of secondary to primary flux. $E_g$ is the location of the spectral break for electrons and $\Delta\gamma _{e^-}$ is the difference in the spectral index before and after the break. The smoothness parameter $s$ governs over which energy range the transition between the two power law indices happens, with higher values of $s$ representing a wider energy range and thus a smoother break in the spectrum. For a hard break \mbox{$(s\to 0)$} the parametrization of the electron flux (eq.~\ref{eq:bplelec}) becomes 
\begin{equation}
 \label{eq:hardbreak}
\begin{split}
\phi_{e^-} &= C_{e^-}E^{\gamma _{e^-} - \Delta\gamma _{e^-}}\, \left(\frac{C_s}{C_{e^-}}\, E^{-\gamma _{e^-}+\gamma _s}\, + e^{-E/E_d}\right) +\, \phi _{\text{extra}}\,\, ; \,\, \text{for}\,\, E \leq E_g\\
 \phi_{e^-} &= C_{e^-}\,\left(\frac{E}{E_g}\right)^{\gamma _{e^-}}\, E_g ^{-\gamma _{e^-}-\Delta\gamma _{e^-}}\, \left(\frac{C_s}{C_{e^-}}\, E^{-\gamma _{e^-}+\gamma _s}\, + e^{-E/E_d}\right) +\, \phi _{\text{extra}}\,\, ;  \,\, \text{for}\,\, E > E_g 
\end{split}
\end{equation}

For $s\rightarrow 1$, the spectral break transitions to the addition of	an electron-only power law spectral component. Thereby, the parametrization can represent both a break from injection or propagation condition, as well as additional contribution from nearby SNRs, with a smooth transition between both scenarios.     

In both cases the primary electron spectrum is cut off exponentially at the energy $E_d$, describing radiative energy loss. The value of $E_d$ may also reflect the contribution from nearby SNR sources and the thickness of the spiral arms as shown in our previous work~\cite{Bhattacharyya:selfcite}. Several fixed values up to 10~TeV were tried and we show results for the 2~TeV and 10~TeV cases as two boundary examples. The results of DAMPE and HESS suggest the existence of a spectral break in the $e^+\, +\, e^-$ spectrum above 1~TeV. 
$E_d=2$~TeV represents a background spectrum compatible with the suggested spectral break, while $E_d=10$~TeV models a smoother flux reduction in the TeV region.


$\gamma _{e^-}-\gamma _s$ is the difference in the spectral indices between primary and secondary CRs, which is fixed to $\gamma _{e^-}-\gamma _s = -\delta$, assuming a diffusion model for the CR propagation with higher energy protons being the source of the secondary electrons and positrons. $\delta$ is the power law index of the rigidity dependence of the spatial diffusion coefficient which is given by 
\begin{equation}
D(R)=D_0\left(\frac{R}{R _0}\right)^{\delta} \, \, ,
\end{equation}    
where $R$ is the rigidity of the particles, $D_0$ is the normalization constant for the spatial diffusion at $R_0$. 
Throughout this work we use $\delta = 0.4$, based on a propagation model tuned to fit the proton spectrum and B/C ratio measurements~\cite{Bhattacharyya:selfcite}, using the GALPROP~\cite{Strong:1998pw, GALPROP_manual} code for the CR propagation calculations.   

Accordingly, the positron flux is given by 
\begin{equation}
 \label{eq:bkposi}
\phi _{e^+} = C_{e^-}E^{\gamma _{e^-}}\left(\frac{C_s}{C_{e^-}}\, E^{-\gamma _{e^-}+\gamma _s}\, \right)\,  +\, \phi _{\text{extra}}\, \, \, ,
\end{equation} 

where $\phi _{\text{extra}}$ is the flux of electron-positron pairs from either DM decay or pulsar source. 





\section{Dark Matter Decay Explanation of the CALET Measurement}\label{sec:DMfit}
To explain the CR positron excess, various models of DM annihilation or decay in the galactic halo have been proposed. In the DM annihilation scenario, a large enhancement of the effective annihilation cross-section, possibly from substructures in the galactic halo~\cite{DMannisubstructure, Kamionkowski:2010mi} or through Sommerfeld or Breit-Wigner enhancement~\cite{Hisano:sommerfeld, Ibe:breitenhancement}, is required to explain the observed relic density and the positron excess simultaneously. DM annihilation models with these enhancements (`boost factor') in the cross-section would exhibit a very large $\gamma$-ray flux from regions with high DM density, such as the Galactic Center and Dwarf Galaxies, resulting in strong constraints on them \cite{Archambault:2017wyh, HESS:2015cda, Cirelli:anniexclude, Bell:anniexclude}.

Such a boost-factor is not required in the decay scenario, as the flux from decay scales with the inverse of the lifetime of the DM, and thus it can more naturally explain the positron excess with a lifetime in the order of \mbox{$\sim 10^{25}\ - \ 10^{26}$s~\cite{Nardi:2008ix, Cirelli:2008pk}}. Among various decay scenarios, DM decaying to three leptons are favored to explain the positron excess, since 3-body decay produces a softer spectrum~\cite{Ibarra:decayofDM, Carone:3Decaysoft} compared to 2-body decay, and the purely leptonic decay products allow for agreement with the recent anti-proton measurements~\cite{Aguilar:antiprotonAMS02}.  TeV scale DM candidates with this decay mode $\left(DM\, \rightarrow\, l\, l\, \nu\right)$ are proposed by the theoretical models described in these Refs.~\cite{Kohri:2009yn,Ibe2013} and the CR signatures from such a decay are discussed in Refs.~\cite{Kohri:2013sva, IBE2015134}.   

In this work we revisit the TeV-scale DM decaying scenario introduced in our previous work~\cite{Bhattacharyya:selfcite}, where the DM decays to a lepton~-~anti-lepton pair and a neutrino $(DM\rightarrow\, l^{\pm} _{i,j,k}\, l^{\mp} _{i,j,k} \, \nu)$, here $i,j,k$ are flavor indices. 
We assume that the Branching Ratios (BR) for the outgoing leptons are free parameters, expressed by the inverse of the decay times $\left(\frac{1}{\tau _e},\, \frac{1}{\tau _{\mu}},\, \frac{1}{\tau _{\tau}}\right)$ for the individual decay channels $(ee\nu,\, \mu\mu\nu ,\, \tau\tau\nu)$. 
We study 3-body leptonic decay of DM as an explanation of the CR positron excess and investigate possible constraints from $\gamma$-ray production. However, we do not consider constraints from direct detection experiments, as interaction of DM with nuclei depends on specific properties of the theoretical model.   
For a fixed DM mass, $\tau$-leptons produce most $\gamma$-rays~\cite{PhysRevD.86.083506} among the leptonic decay products, and the $\gamma$-ray emission can be significantly reduced if the decay is constrained to only $\mu\mu\nu$ and $ee\nu$ channels~\cite{Bhattacharyya:selfcite}. Therefore in this work we consider decay excluding the $\tau\tau\nu$ channel when fitting DM decay models to the measurements. 

For the DM distribution in the galactic halo, we assume a Navarro-Frenk-White (NFW) profile~\cite{Navarro:1996gj} which is given by
\begin{equation}
 \label{eq:NFWprofile}
\rho = \frac{\delta _c \rho _c}{(r/r_s)(1+r/r_s)^2}\,\, ,
\end{equation}
with $\delta _c$ defined as  
\begin{equation}
 \label{eq:nfw}
\delta _c = \frac{200}{3}\frac{c_v ^3}{\text{ln}(1+c_v)-(c_v/(1+c_v))} \,\, ,
\end{equation} 
where $c_v$ is the ratio of virial radius $(r_v)$ and scale radius $(r_s)$, and we assume \mbox{$c_v=10$~\cite{Lokas:2000mu}}. $\rho _c$ is determined from the mass of the halo as 
\begin{equation}
\rho _c = \frac{\frac{4}{3}\pi r_v ^3}{M_v} \,\, ,
\end{equation}
where $r_v,\,M_v$ are taken as $200$ kpc and $1.5\times 10^{12}\text{M}_{\odot}$~\cite{Dehnen:2006cm}.
With these assumptions, we calculate the injected particles per volume and time in the decaying DM scenario using the equation 
\begin{equation}
Q = \Gamma \frac{\rho}{\text{M}_{\text{DM}}}\frac{dN}{dE} \, .
\end{equation} 

Since the the flux of electron-positron CRs from the DM decay scales with the inverse of the decay time of outgoing leptons, it can be written as
\begin{equation}
 \label{eq:DMflux}
\phi _{\text{DM}} = \frac{1}{\tau _e}\, \phi _e \, +\, \frac{1}{\tau _{\mu}}\, \phi _{\mu}\, +\, \frac{1}{\tau _{\tau}}\, \phi _{\tau}\,  , 
\end{equation}
with $\phi _e,\,\phi _{\mu},\, \phi _{\tau}$ being the $e^+$ (identical to $e^-$) decay spectra for $ee\nu,\,\mu\mu\nu,\, \tau\tau\nu$ channel, respectively, calculated with PYTHIA~\cite{Sjostrand:2007gs} and propagated using GALPROP. The propagation parameters are optimized to reflect the current proton and B/C ratio data as described in our previous work~\cite{Bhattacharyya:selfcite}.

\subsection{Parametrization-Fit to Current Measurements with Dark Matter as Extra Source}
Using the background parametrization (eq.~\ref{eq:bplelec}) and DM decay as the extra source, we do a combined $\chi ^2$ fit to the CALET $e^+\,+\,e^-$ spectrum and the AMS-02 $e^+$ spectrum to determine the best fit parameters. The $\chi ^2$ value is given by 
\begin{equation}
\chi ^2 = \sum _i \frac {\left(\phi _{\text{obs},\, i}-\phi _{\text{model},\, i}\right)^2}{\sigma _{i} ^2}\, \, , 
\end{equation}
where $\phi _{\text{obs},\, i}$ and $\sigma _i $ are respectively flux and errors of the experimental data for the $i-$th energy bin and $i$ runs over the number of data points.  
The energy range of the data points used in this fitting is from 10~GeV to 3~TeV. The upper bound of the fit range is given by the highest energy data point from the CALET measurement. The data points of both CALET and AMS-02 measurements below 10~GeV are not used, since the flux of the CRs in this energy range is strongly influenced by solar modulation. The CALET and AMS-02 data sets were taken during different periods of the solar cycle, and at energies $\leq 10$~GeV, solar modulation can be modeled including time and charge dependence~\cite{Cholis:2015gna}. Therefore, for electrons and positrons solar modulation potential would have to be modeled separately, introducing several new free parameters. We assume 
a solar modulation model based on the force field approximation~\cite{Cholis:2015gna}, in which the charge dependent effects are represented by an increased solar modulation potential for one charge sign at low energy, while particles above 10~GeV are only affected by the generally lower common potential $(\phi)$, which we assume as 0.5~GV. The predicted range of the common potential is about 0.4~GV$\, \textendash \, $0.6~GV~\cite{Cholis:2015gna, Corti:2015bqi} and we show results of the fits with $\phi=0.4$~GV and $\phi=0.6$~GV for comparison. While the value of $\phi$ is kept fixed in the fitting process, the location of the spectral break, $E_g$ is taken as a free parameter. The fit quality depends on the smoothness term $(s)$, for which several fixed values from $s=0$ to $s=1$, with a step size of 0.05, were used.    
Several fixed values for the DM mass in the range from 600~GeV to 4~TeV were fitted, determining the range in which the model is allowed based on the combined $\chi ^2$ of the fit to CALET $e^+\,+\,e^-$ and AMS-02 $e^+$ spectrum in comparison to the $95\%$ CL threshold.

\begin{figure}[htp]
\includegraphics[clip,width=0.77\columnwidth]{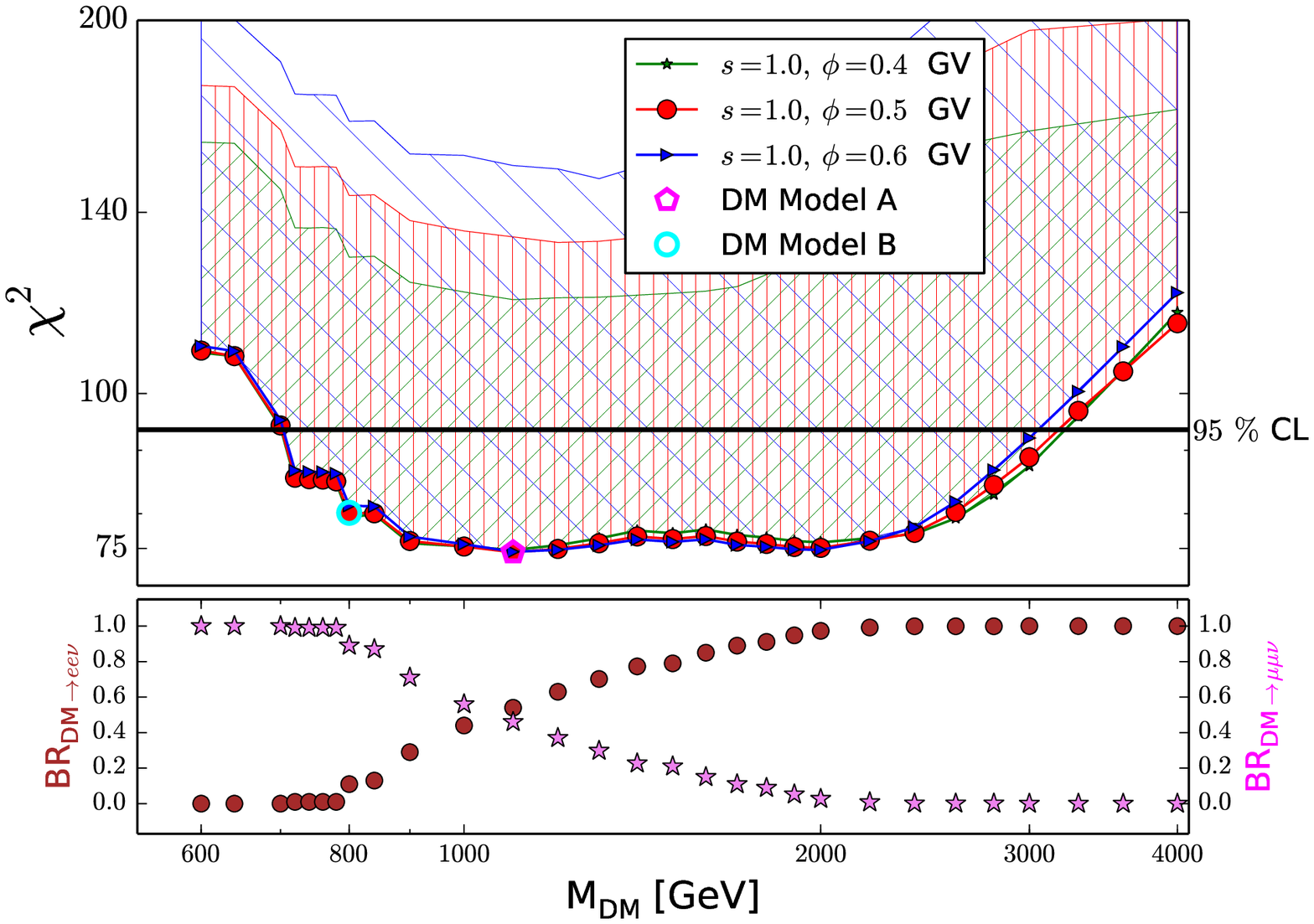} (a) \\
\includegraphics[clip,width=0.77\columnwidth]{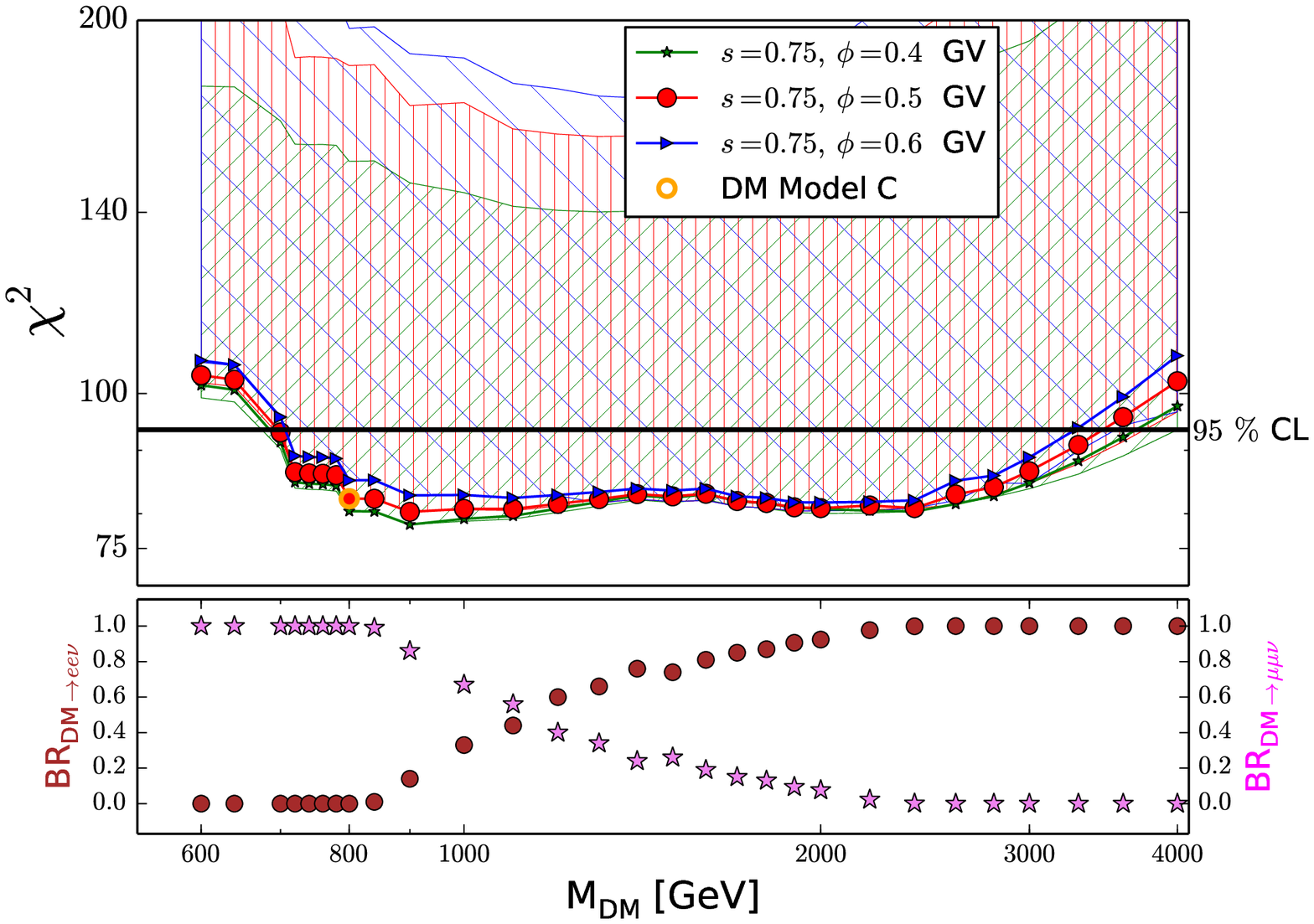} (b) %
\caption{Figure (a): Upper panel shows the dependence of minimum $\chi ^2$ obtained from the  combined fit to the $e^+\, +\, e^-$ flux (CALET) and $e^+$ flux (AMS-02) for fixed values of $s=1.0$ on the DM mass. Green, red and blue line represent solar modulation potential $\phi =0.4$ GV, $\phi =0.5$~GV and $\phi=0.6$ GV, respectively and the shaded regions with same colors depict the minimum $\chi ^2$ obtained using different values of smoothness $(s)$ for the studied values of $\phi$. DM Model A and \mbox{DM Model B} are marked with magenta and cyan dot respectively. Black horizontal line denotes the $95\%\,$ CL. Lower panel shows the variation of branching ratios from the best-fit cases for $DM \rightarrow ee\nu$ channel (pink stars) and $DM \rightarrow \mu\mu\nu$ (brown circle) with DM mass for $s=1.0,\, \phi =0.5$~GV. Figure (b): Same as figure (a), but now the results are shown for background cut-off energy $(E_d)$ of 2~TeV, with $s=0.75$. DM Model C is denoted by yellow dot. \label{fig:DMmassandchi}}
\end{figure}

We do separate studies for background cut-off energy $E_d = 10$~TeV and \mbox{$E_d = 2$ TeV}, with the goodness of fit as a function of the DM mass and the allowed range of DM masses being presented in figure~\ref{fig:DMmassandchi}. The shaded region shows the range of $\chi ^2$ values obtained with different smoothness for each DM mass, for a fixed value of the solar modulation potential $\phi$. It is shown that the minimum $\chi ^2$ curves are only slightly shifted by the choice of solar modulation potential $\phi$. 
The allowed range of DM mass from the fit to measurements with respect to a $95\%$~CL threshold (93.94 for 73 degrees of freedom) is nearly the same for both values of $E_d$, ranging from $\sim 700$~GeV to $\sim 3$~TeV, showing that the position of the background spectrum drop-off also does not influence this result significantly.

The overall minimum $\chi ^2$ value of 74.5 (with 73 degrees of freedom) for the DM decay model is obtained for a DM mass of 1.1~TeV with fixed values of $s = 1.0$ and $E_d =10$~TeV and the other characteristic fit parameter values are given in Table~\ref{Table:ModelCALETfitDM}. From now onwards we denote this best fit case as DM Model A. The best fit case for DM Model A is shown in figure~\ref{fig:800GeVDMEd2T}, with branching ratios of $46\%$ for the $ee\nu$ channel and $54\%$ for the $\mu\mu\nu$ channel, with the lifetime of the DM being $3.03\times 10^{26}$~s. 

\begin{table}
\tbl{Parameters obtained from the best fit of different studied models to combined CALET $e^+\, +\, e^-$ and AMS-02 $e^+$ spectra is listed here. Upper line (lower line) of each cell shows the values obtained without (with) including systematic error of CALET data. \label{Table:ModelCALETfitDM}}{
		\tabcolsep=0.05cm
		\begin{tabular}{|l|c|c|c|c|c|c|c|c|c|c|}\hline
		\makecell{\textbf{Model}\\ DM}&\makecell{$C_{e^-}$\\$\left(\frac{\text{\tiny{GeV}}}{\text{s}\,\, \text{m}^{2}\,\, \text{sr}}\right)$}&$\frac{C_s}{C_{e^-}}$& $\gamma _{e^-}$& $\Delta\gamma _{e^-}$&\makecell{$E_g$\\ GeV}& $s$&\makecell{$\tau$\\\tiny{$\times 10^{26}$ s}}&$BR_{el}$&$BR_{\mu}$&$\chi^2$ \\	
		\multirow{6}{*}{\makecell{Model A\\$M=1.1$ TeV \\ $E_d=10$~TeV}}&\multirow{4}{*}{$859$}&\multirow{4}{*}{$0.034$}&\multirow{4}{*}{$2.90$}&\multirow{4}{*}{$0.681$}&\multirow{4}{*}{$87.6$}&\multirow{4}{*}{$1.0$}&\multirow{4}{*}{$3.03$}&\multirow{4}{*}{$0.46$}&\multirow{4}{*}{0.54}&\multirow{4}{*}{$74.49$}\\\cline{1-11}
		&&&&&&&&&&\\
		&\multirow{4}{*}{$863$}&\multirow{4}{*}{$0.036$}&\multirow{4}{*}{$2.91$}&\multirow{4}{*}{$0.679$}&\multirow{4}{*}{$87.6$}&\multirow{4}{*}{$1.0$}&\multirow{4}{*}{$3.04$}&\multirow{4}{*}{$0.47$}&\multirow{4}{*}{0.53}&\multirow{4}{*}{$29.83$}\\\cline{2-11}
		&&&&&&&&&&\\
		&&&&&&&&&&\\\hline
		\multirow{4}{*}{\makecell{Model B\\$M=0.8$ TeV \\ $E_d=10$ TeV}}&\multirow{2}{*}{$846$}&\multirow{2}{*}{$0.031$}&\multirow{2}{*}{$2.84$}&\multirow{2}{*}{$0.713$}&\multirow{2}{*}{$115.3$}&\multirow{2}{*}{$1.0$}&\multirow{2}{*}{$5.85$}&\multirow{2}{*}{$0.89$}&\multirow{2}{*}{0.11}&\multirow{2}{*}{80.43}\\
		&\multirow{4}{*}{$856$}&\multirow{4}{*}{$0.031$}&\multirow{4}{*}{$2.84$}&\multirow{4}{*}{$0.72$}&\multirow{4}{*}{$115.3$}&\multirow{4}{*}{$1.0$}&\multirow{4}{*}{$6.13$}&\multirow{4}{*}{$0.94$}&\multirow{4}{*}{0.06}&\multirow{4}{*}{$35.21$}\\\cline{2-11}	
		&&&&&&&&&&\\
		&&&&&&&&&&\\\hline
		\multirow{4}{*}{\makecell{Model C\\ $M=0.8$ TeV\\ $E_d=2$ TeV}}&\multirow{2}{*}{$835$}&\multirow{2}{*}{$0.027$}&\multirow{2}{*}{$2.78$}&\multirow{2}{*}{$0.725$}&\multirow{2}{*}{$107.1$}&\multirow{2}{*}{$0.75$}&\multirow{2}{*}{$7.40$}&\multirow{2}{*}{$1.0$}&\multirow{2}{*}{0.0}&\multirow{2}{*}{$82.50$}\\
		&\multirow{4}{*}{$838$}&\multirow{4}{*}{$0.027$}&\multirow{4}{*}{$2.79$}&\multirow{4}{*}{$0.723$}&\multirow{4}{*}{$106.9$}&\multirow{4}{*}{$0.75$}&\multirow{4}{*}{$7.25$}&\multirow{4}{*}{$1.0$}&\multirow{4}{*}{0.0}&\multirow{4}{*}{$37.02$}\\\cline{2-11}
		&&&&&&&&&&\\
		&&&&&&&&&&\\\cline{1-11}
		\end{tabular}}
	\end{table}

The $e^+\,+\,e^-$ spectrum as measured by CALET features a step-like structure around 400~GeV, which is reflected in the $\chi ^2$ values and best fitting branching ratios in the mass range from 720~GeV to 800~GeV. As shown on the lower panels of each plot in figure~\ref{fig:DMmassandchi}, the decay of DM with mass $\sim 800$~GeV and a high branching fraction for the electron channel can reproduce this step-like structure in the \mbox{$e^+\,+\,e^-$} spectrum around 400~GeV, causing a clear reduction in $\chi ^2$ compared to lower masses, which makes 800~GeV DM especially interesting to study. With further increasing DM mass, the contribution of the softer spectrum from the $\mu\mu\nu$ channel increases, resulting in a relatively smoother step which also models the spectrum well as depicted by the further decreasing $\chi ^2$ up to 1.1~TeV. 

For 800~GeV DM with $E_d$ set to 10~TeV, the branching ratio obtained are $11\%$ for the $\mu\mu\nu$ channel and $89\%$ for the $ee\nu$ channel with the lifetime of the DM being $5.85\times 10^{26}$~s, which we will call DM Model B. 
With $E_d=2$~TeV, the branching ratios obtained are $100\%$ for the $ee\nu$ channel and no contribution from $\mu\mu\nu$ channel, with lifetime of DM being $7.40\times 10^{26}$~s, which we will call DM Model C. The resulting fit plots for DM Model B and DM Model C are also shown in figure~\ref{fig:800GeVDMEd2T}, demonstrating the correspondence of the step-like structure with the high-energy end of the DM source spectrum. 

Our choices of data sample, fixed background parameters and solar modulation model introduce a bias in the models obtained from the fitting, making this not an exhaustive search for all possibilities to explain the positron excess with the chosen Dark Matter decay scenario. The three models are selected as examples for the existence of possibly viable models.      

The parameter values obtained from the fit are listed in Table~\ref{Table:ModelCALETfitDM}, with the value of the diffuse primary $e^-$ flux coefficient $(C_{e^-})$ given for an energy of 1~GeV. The $\frac{C_s}{C_{e^-}}$ parameter represents a scaling of the secondary flux, corresponding to uncertainties on the density of the primary particles in the ISM, choice of propagation parameters, and the differential cross-section for the spallation process~\cite{DiMauro:2015jxa}. The values obtained for this parameter and the secondary spectrum index (which is tied to the higher energy primary index as $\gamma _s = \gamma _{e^-} + 0.4$) from the fit of the different extra source models to CALET data all yield a secondary flux component which is comparable with the results derived in these Refs.~\cite{DiMauro:2015jxa, DiMauro:2014iia, DiMauro:2017jpu} from primary nuclei measurements.

\begin{landscape}\centering
 \begin{figure}\centering
 \begin{minipage}[t]{.3\textwidth}
 \centering
 \vspace{2.7cm}
 \includegraphics[width=65mm, height=84mm]{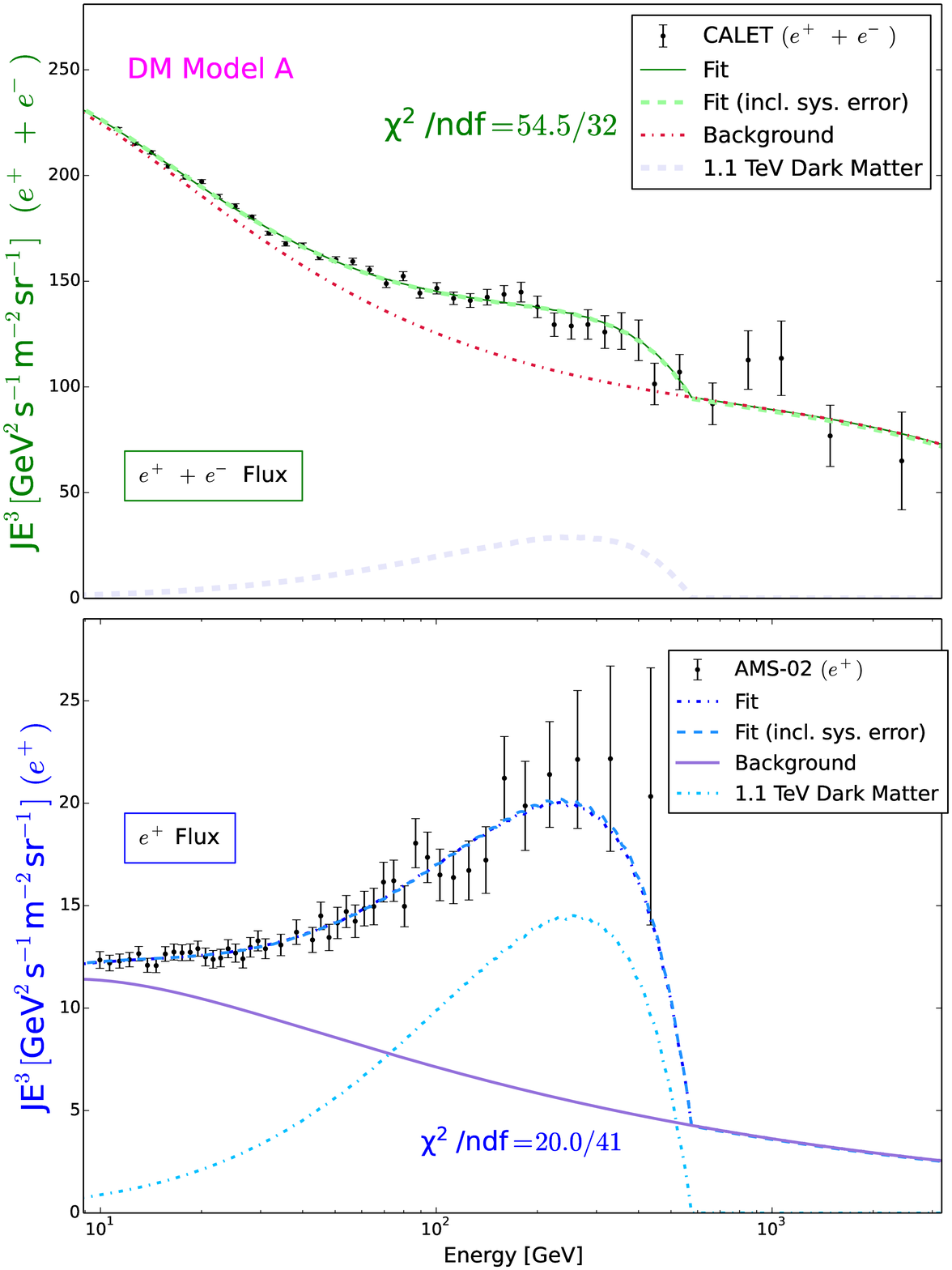} (a)
 \end{minipage}
 \begin{minipage}[t]{.3\textwidth}
 \centering
 \vspace{2.7cm}
 \includegraphics[width=65mm, height=84mm]{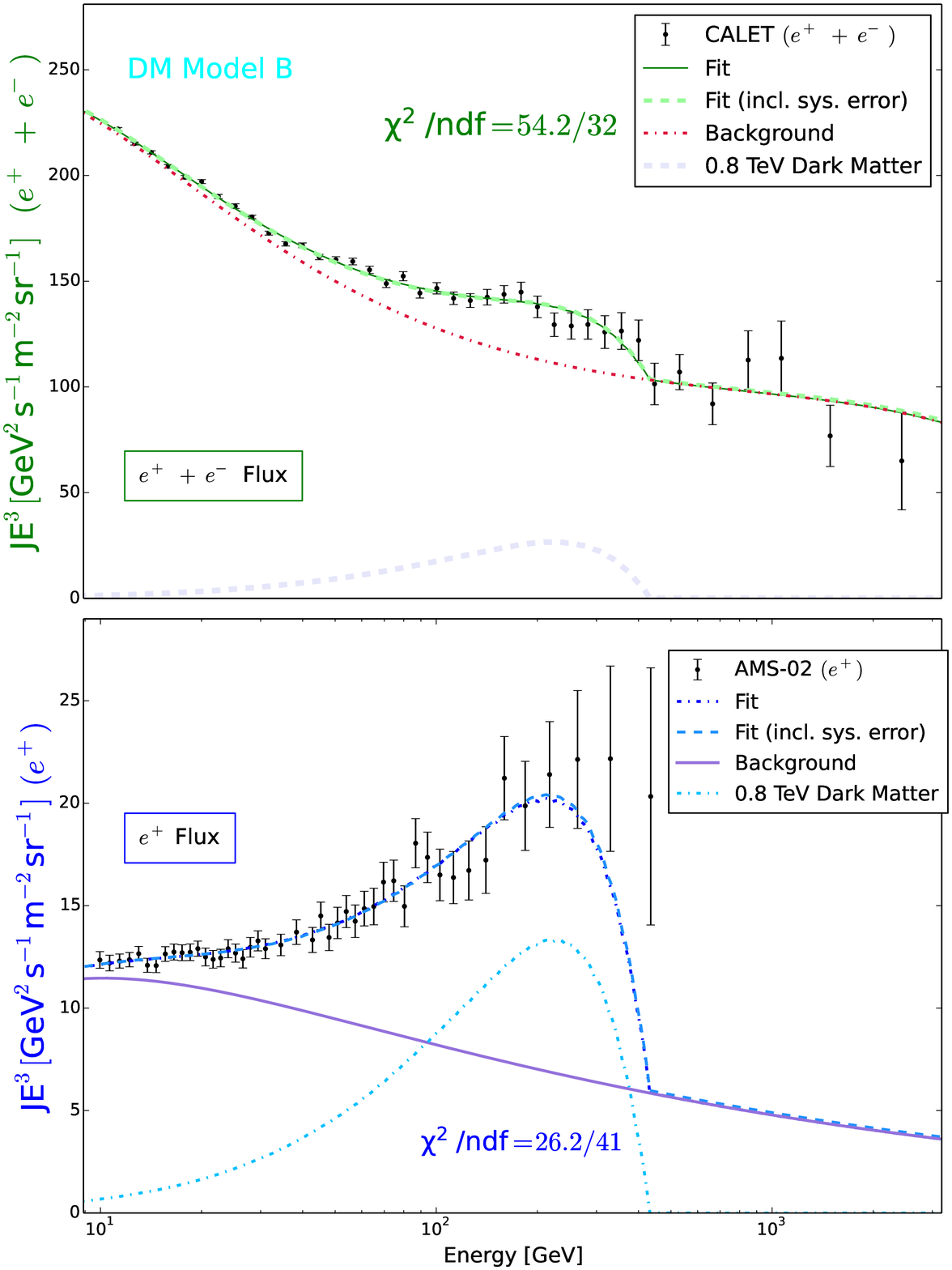} (b)
 \end{minipage}
 \begin{minipage}[t]{.3\textwidth}
 \centering
 \vspace{2.7cm}
 \includegraphics[width=66mm, height=84mm]{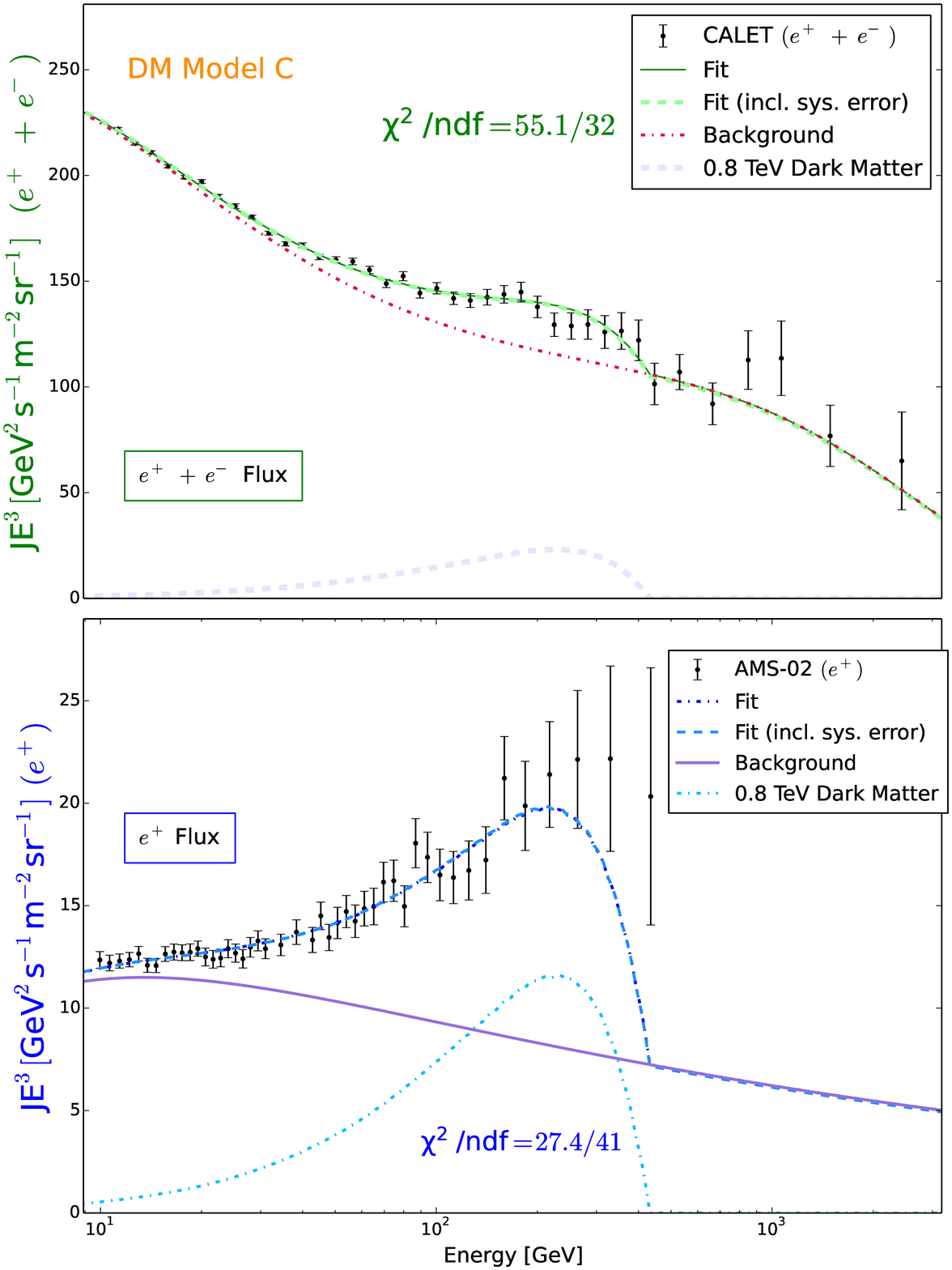} (c)
 \caption{Figure (a): Broken power law (eq.~\ref{eq:bplelec}) with DM Model A as extra source is taken as a test model (green line) to describe $e^+\, +\, e^-$ spectrum measured by CALET (upper panel) and $e^+$ flux measured by AMS-02 (black error bars) in the lower panel. The red-dotted line represents the background $e^+\, +\, e^-$ spectrum. Grey dashed lines in upper panel and blue dotted lines in lower panel show the contribution from the DM to the $e^+\, +\, e^-$ and $e^+$ spectrum, respectively. The best fit obtained including systematic errors is presented with green and blue dashed lines in the upper and lower panels, respectively. Similarly, the fit plots for DM Model B and DM Model C are shown in figure (b) and figure (c), respectively. The fit parameters are listed in Table~\ref{Table:ModelCALETfitDM}. \label{fig:800GeVDMEd2T}}
 \end{minipage}
 \vfill
 \end{figure}
\end{landscape}

For the CALET measurement, only the statistical error is taken into account in this fitting, as a significant part of the systematic errors quoted in Ref.\cite{PhysRevLett:CALET} is expected not to be energy dependent or exhibit a smooth energy dependence, which would be compensated by the variability of the parametrization through the normalization and power law index parameters. In regard to this, we examined the effect of the systematic errors on the obtained fit parameter values, which are listed in Table~\ref{Table:ModelCALETfitDM}, showing that there is no significant shift from those without systematic errors.

\subsection{Diffuse $\gamma$-ray Constraints on Dark Matter Decay}

Decay of DM particles in the galactic halo would contribute to the diffuse $\gamma$-ray flux. The leptonic decay products from the fermionic DM model should produce $\gamma$-rays via Final State Radiation (FSR) and also the interaction of the charged decay products with the ISM should produce secondary $\gamma$-rays via Inverse Compton and Bremsstrahlung processes. Fermi-LAT measured the total extra-galactic $\gamma$-ray background (EGB) at higher latitudes $\left(\left|b\right|>20^{\degree}\right)$, which is the sum of the isotropic diffuse gamma ray (IGRB) and flux from the resolved sources~\cite{Ackermann:2014usa}. Due to the measurement at higher latitudes, the background from astrophysical sources is strongly reduced compared to the galactic plane. This measurement constrains DM decay in the galactic halo, which should contribute to the IGRB. The residual contribution from astrophysical sources depends on the different modelings of the diffuse galactic emission~\cite{Ackermann:2014usa, DiMauro:2015tfa} but the total measured $\gamma$-ray flux can be considered a conservative bound for models of DM decay. 
To calculate the primary $\gamma$-ray flux from FSR and decay of primary charged particles, we used the PYTHIA~\cite{Sjostrand:2007gs} event generator, and numerically integrate the yield over the assumed NFW halo profile. The flux of secondary $\gamma$-rays from the interaction of the charged decay products with the ISM is calculated with GALPROP assuming the default Interstellar Radiation Field (ISRF) distribution~\cite{Porter:2005qx}. Also, the extragalactic flux from DM decay should contribute to the IGRB flux. Following Ref.~\cite{Ando:2015qda} we calculated the primary extra-galactic flux from DM decay as, 

\begin{equation}
\begin{split}
E^2 _{\gamma}\, \phi_{\gamma}^{\text{DM}} & = 1.4\, \times\, 10^{-7}\text{GeV}\, \text{cm}^{-2}\, \text{s}^{-1}\, \text{sr}^{-1} \left(\frac{1\, \text{TeV}}{m_{\text{DM}}}\right)\, \left(\frac{10^{27}\, \text{s}}{\tau _{\text{DM}}}\right)\, \left(\frac{E_{\gamma}}{100\, \text{GeV}}\right) \\
& \times \int _{E_{\gamma}}^{\infty} \,  \frac{e^{-\tau _{\text{od}}}\,\, { \frac{dN}{dE'_{\gamma}} }}{\sqrt{\Omega _{\Lambda}+\Omega _{\text{m}}(E'_{\gamma}/E_{\gamma})^3}}\,  dE'_{\gamma}\, \, \, ,
\end{split}
\end{equation}
where $\tau _{\text{od}}$ is optical depth and we assume the conservative case of a completely transparent universe $\left(\tau _{\text{od}}=0\right)$. The matter and dark energy density $(\Omega _m, \Omega _{\Lambda})$ are taken as 0.315 and 0.685 respectively~\cite{Ade:2015xua}.  

\begin{figure}
 \centering 
 \includegraphics[width=1.1\linewidth]{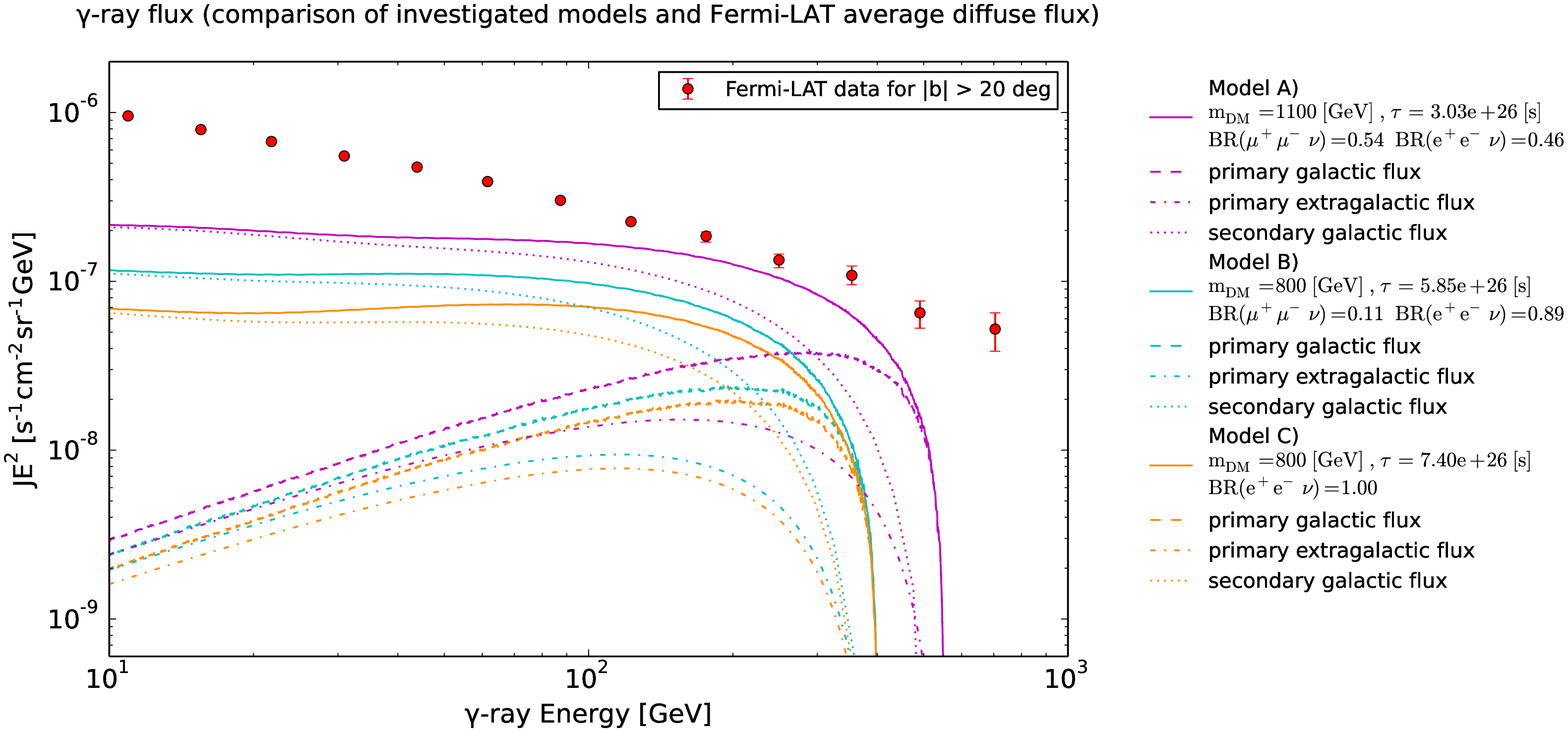} (a)\\
 \includegraphics[width=0.49\linewidth]{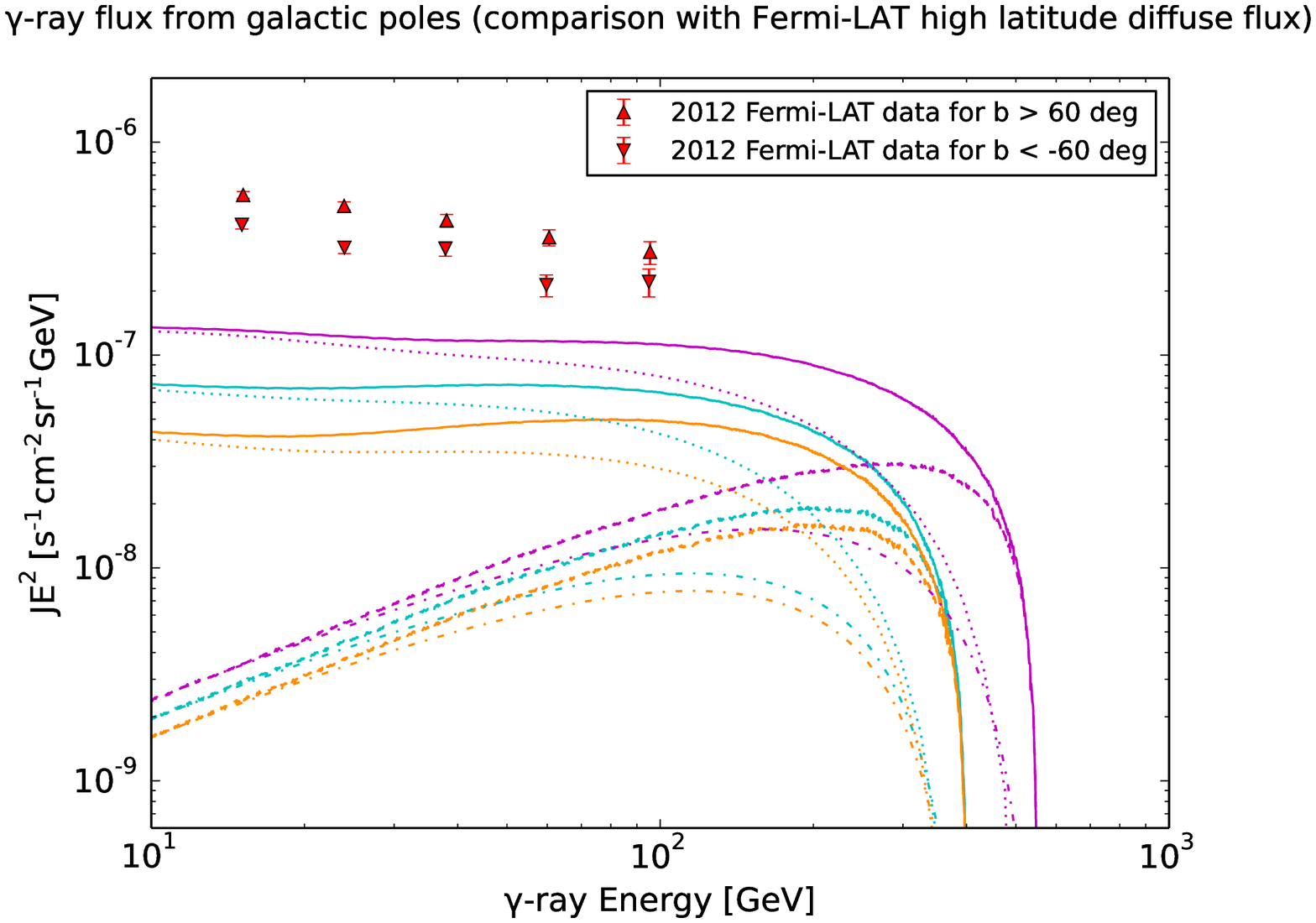}
 \includegraphics[width=0.49\linewidth]{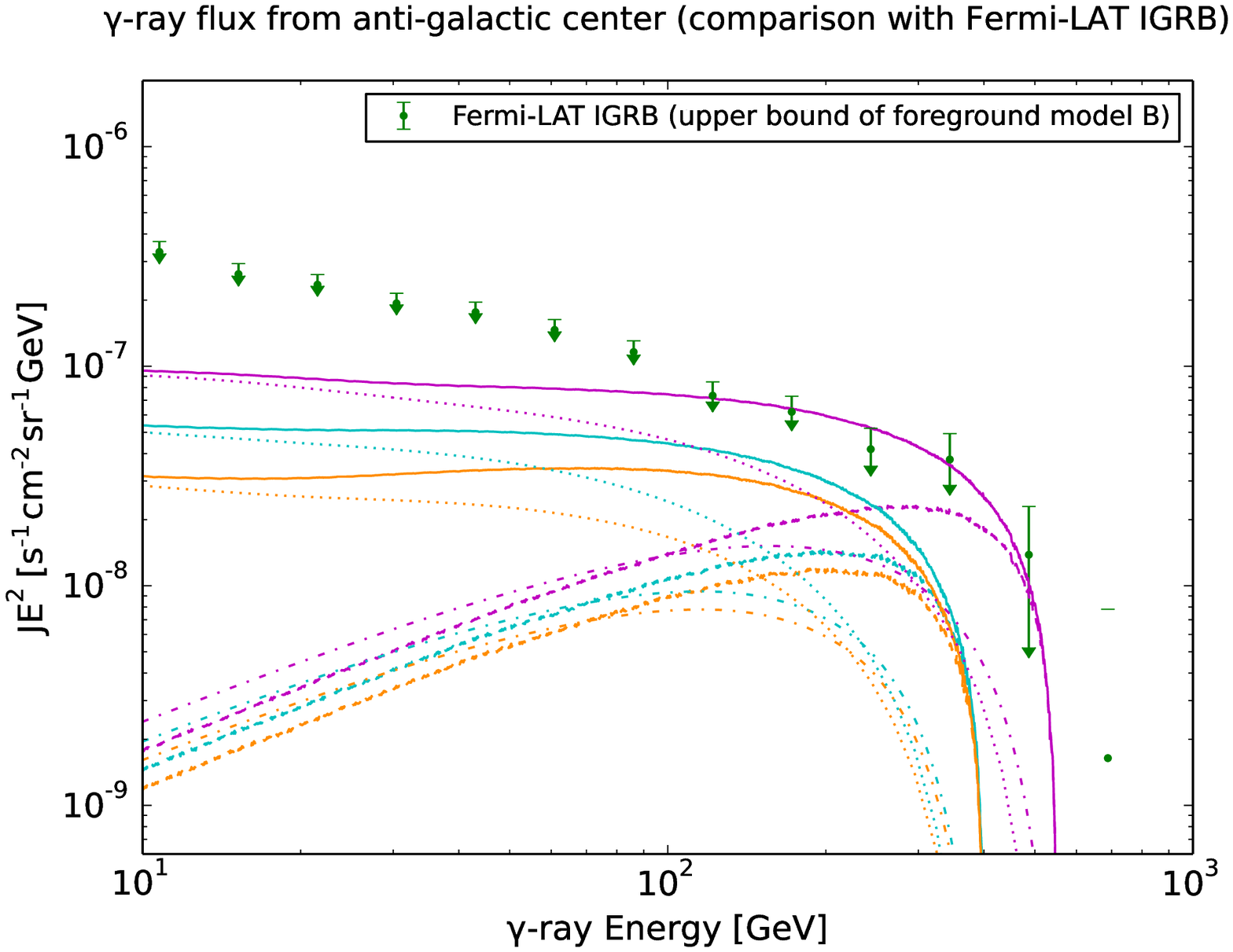}	
 \caption{Upper Panel: Primary and secondary $\gamma$-ray flux accompanied by DM decay to $ee\nu$ and $\mu\mu\nu$ channel for DM Model A, DM Model B and DM Model C are shown with magenta, cyan and orange lines, respectively. The  dashed lines show the primary components, dotted lines show the secondary components and dash-dot lines show the extragalactic components for each DM mass. Lower Panel: The comparison of $\gamma$-ray flux from the DM models with the high latitude $\left(\left|b\right|>60^{\degree}\right)$ results from Fermi-LAT~\protect\cite{Ackermann:2012pya} (left) is shown. The $\gamma$-ray flux at the anti-galactic center from the DM models are compared with the Fermi-LAT IGRB measurement for foreground model B (right). Representation for the different lines are same as in figure~\ref{fig:gammafromDM}. \label{fig:gammafromDM}}
\end{figure}

The $\gamma$-ray flux obtained with these calculations for the DM Model A is plotted in figure~\ref{fig:gammafromDM}.  
This is compared with the scenario of 800~GeV DM decay, for which the total $\chi ^2$ from the fit to CALET $e^+\, +\, e^-$ flux and AMS-02 $e^+$ flux is also well below the $95\%$~CL threshold  for $E_d=10$~TeV and for $E_d=2$~TeV, as shown in figure~\ref{fig:DMmassandchi}. \mbox{DM Model C} with $100\%$ branching to $ee\nu$ channel has the lowest $\gamma$-ray flux.

The spectrum of diffuse $\gamma$-ray flux up to 820 GeV is published by the \mbox{Fermi-LAT} collaboration averaged over all directions with latitude $\left|b\right|>20^{\degree}$. As both the flux from DM decay and the measured diffuse flux show anisotropy, we compare them in two additional ways: The minimum of the overall measured flux is found at high latitude, and we compare the flux up to 100~GeV as published by \mbox{Fermi-LAT} in 2012~\cite{Ackermann:2012pya} for $\left|b\right|>60^{\degree}$ with the DM flux calculated for the same sky region. As shown in the lower left panel of figure~\ref{fig:gammafromDM}, the flux from the DM decay expected from this region is well below the measured flux. Furthermore, we compare the minimum of the expected flux from DM decay in direction of the anti-galactic center with the \mbox{Fermi-LAT} IGRB results, which is the modeled isotropic component of the flux after the subtraction of the galactic foreground.  The primary galactic and extra-galactic flux towards the anti-galactic center are found to be of the same order of magnitude below 100~GeV, in line with the results given in Ref.~\cite{Cirelli:2012ut}. As shown in the lower right panel of figure~\ref{fig:gammafromDM}, the flux from the DM decay for all three models is compatible with the IGRB calculated with \mbox{foreground model B} in Ref.~\cite{Ackermann:2014usa}, the model yielding the highest IGRB flux.

\section{Single Pulsar Explanation of the CALET Measurement}\label{sec:pulfit}

As parametrization of the pulsar extra source, we use an exponentially cut-off power-law spectrum, which corresponds to a single young pulsar~\cite{Motz:2015cua}. It is shown in Ref.~\cite{Feng:2015uta} that several pulsars from the ATNF catalog~\cite{Manchester:2004bp} which are at a \mbox{distance $<0.5$~kpc} from the  Earth and have an age \mbox{$\approx 0.45\,\sim\, 4.5\times 10^5$ years} could each explain the AMS-02 positron excess. While a recent publication of the High-Altitude Water Cherenkov (HAWC) collaboration questions the validity of the nearby pulsar explanation~\cite{Abeysekara:2017old} based on the low diffusion coefficient derived from their measurements at several 10~TeV energies, we would like to point out that this conclusion requires that the diffusion coefficient measured near the sources is universal throughout the galaxy, and also requires the extrapolation  of this coefficient by the assumption of a power law to the $\sim \, 100$~GeV region relevant for the sources of the positron excess.


Thus we adopt the single pulsar model as a generic scenario explaining positron excess in comparison to the decaying DM model. The pulsar extra source flux is determined by the normalization factor $C_{\text{pwn}}$, the power-law index $\gamma _{\text{pwn}}$ and the cut-off energy $E_{\text{pwn}}$, and can be written as  
  \begin{equation}
   \label{eq:extrasourcepulsar}
\phi _{\text{pwn}} = C_{\text{pwn}}E^{\gamma _{\text{pwn}}}e^{-\left(\frac{E}{E_{\text{pwn}}}\right)} \, \, . 
  \end{equation}
The same values of the smoothness parameter~$(s)$ as in the decaying DM scenario were tested. In addition to the chosen value of $0.5$~GV for the solar modulation  potential, we also studied $\phi=0.4$~GV and $\phi=0.6$~GV.   

The combined $e^+\, +\, e^-$ flux and $e^+$ flux can be fitted with $\chi ^2$ below $95\, \%$~CL using this pulsar model, also if the break in the primary electron spectrum is hard (eq.~\ref{eq:hardbreak}). 
In the DM decay scenario on the other hand, for hard break no values of DM mass in the range (600~GeV to 4~TeV) can explain the measurements at $95\%$~CL for the tested fixed values of solar modulation potential, showing the necessity of a smooth break.

\begin{figure}[htp]
\includegraphics[clip,width=0.8\columnwidth]{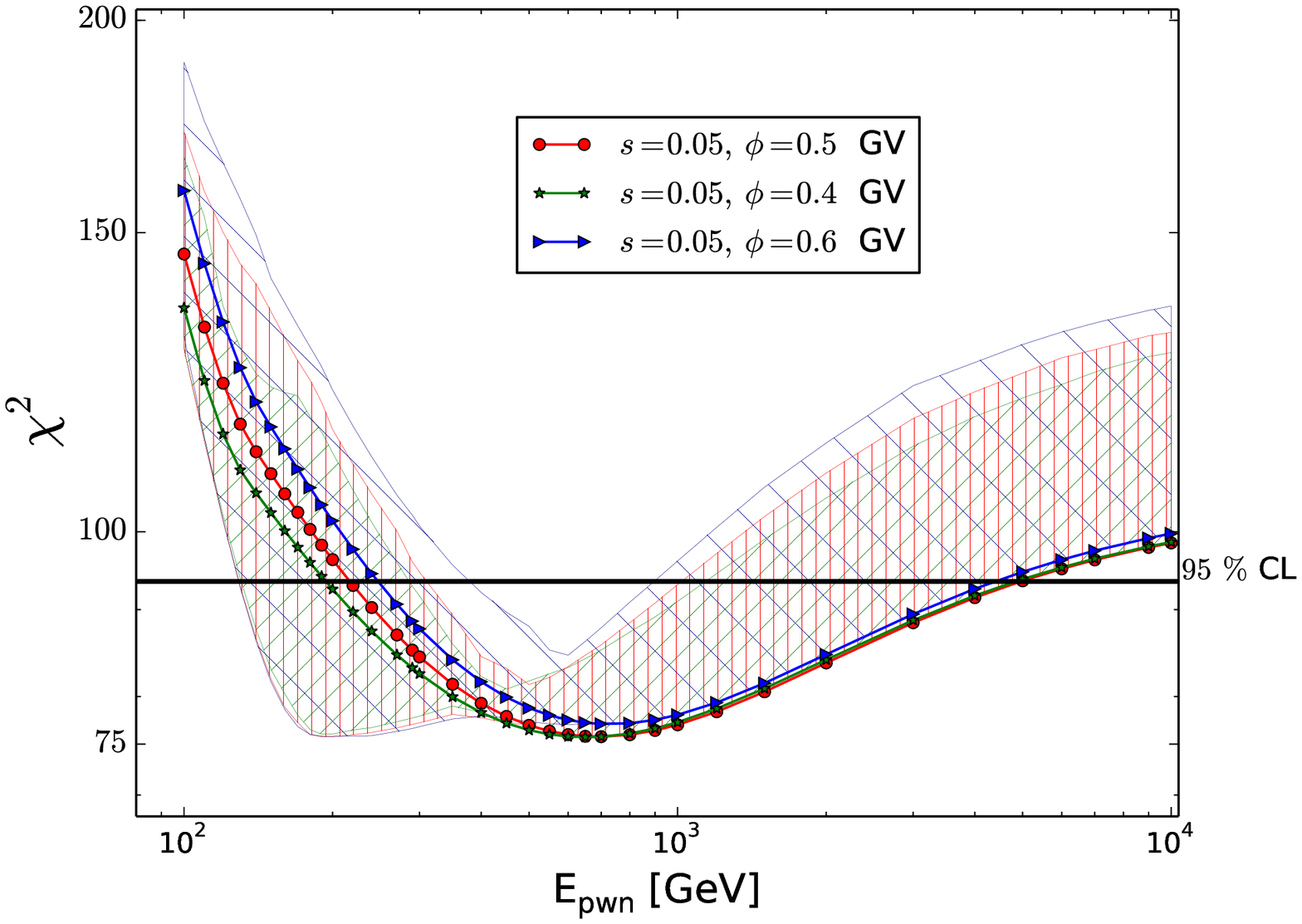} (a) \\
\includegraphics[clip,width=0.8\columnwidth]{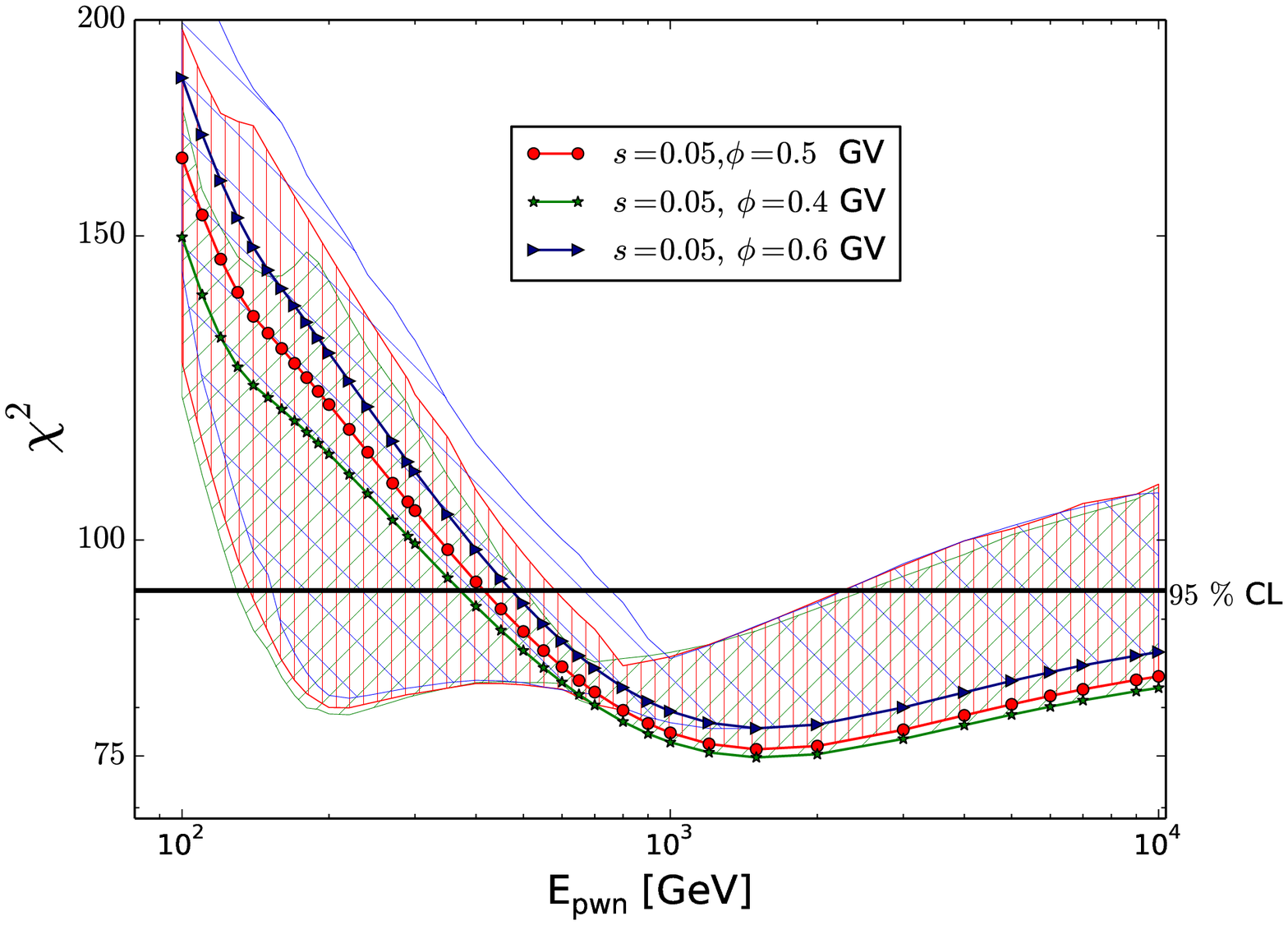} (b)%
\caption{Figure shows the dependence of minimum $\chi ^2$ obtained from the  combined fit to the $e^+\, +\, e^-$ flux (CALET) and $e^+$ flux (AMS-02) for fixed values of $s=0.05$ on the pulsar cut-off energy $(E_{\text{pwn}})$. Green, red and blue line represent solar modulation potential $\phi =0.4$ GV, $\phi =0.5$~GV and $\phi=0.6$ GV, respectively and the shaded regions with same colors depict the minimum $\chi ^2$ obtained using different values of smoothness $(s)$ for the studied values of $\phi$. In the upper panel (a) the results are shown for $E_d=10$~TeV and in the lower panel (b) they are for $E_d=2$~TeV and the minimum $\chi ^2$ is obtained for $s=0.05$. \label{fig:pulcutvschi}}
\end{figure}

The allowed range for the pulsar-cut off energy $(E_{\text{pwn}})$ compared to the $95\, \%$~CL threshold from the combined fit to the CALET $e^+\, +\, e^-$ flux and the AMS-02 $e^+$ flux are shown in figure~\ref{fig:pulcutvschi} for $E_d = 10$~TeV and $E_d = 2$~TeV. The shaded regions in both figures show the minimum value of $\chi ^2$ for different smoothness from the hard break case to $s=1$, each region representing one fixed value of solar modulation potential. Pulsar cut-off energies larger than $\sim \, 200$~GeV\textendash $400$~GeV, and depending on $E_d$ and solar modulation potential up to $\sim 5$~TeV or more, are found to be allowed at $95\%$~CL.

\begin{figure}
 \includegraphics[width=0.9\textwidth]{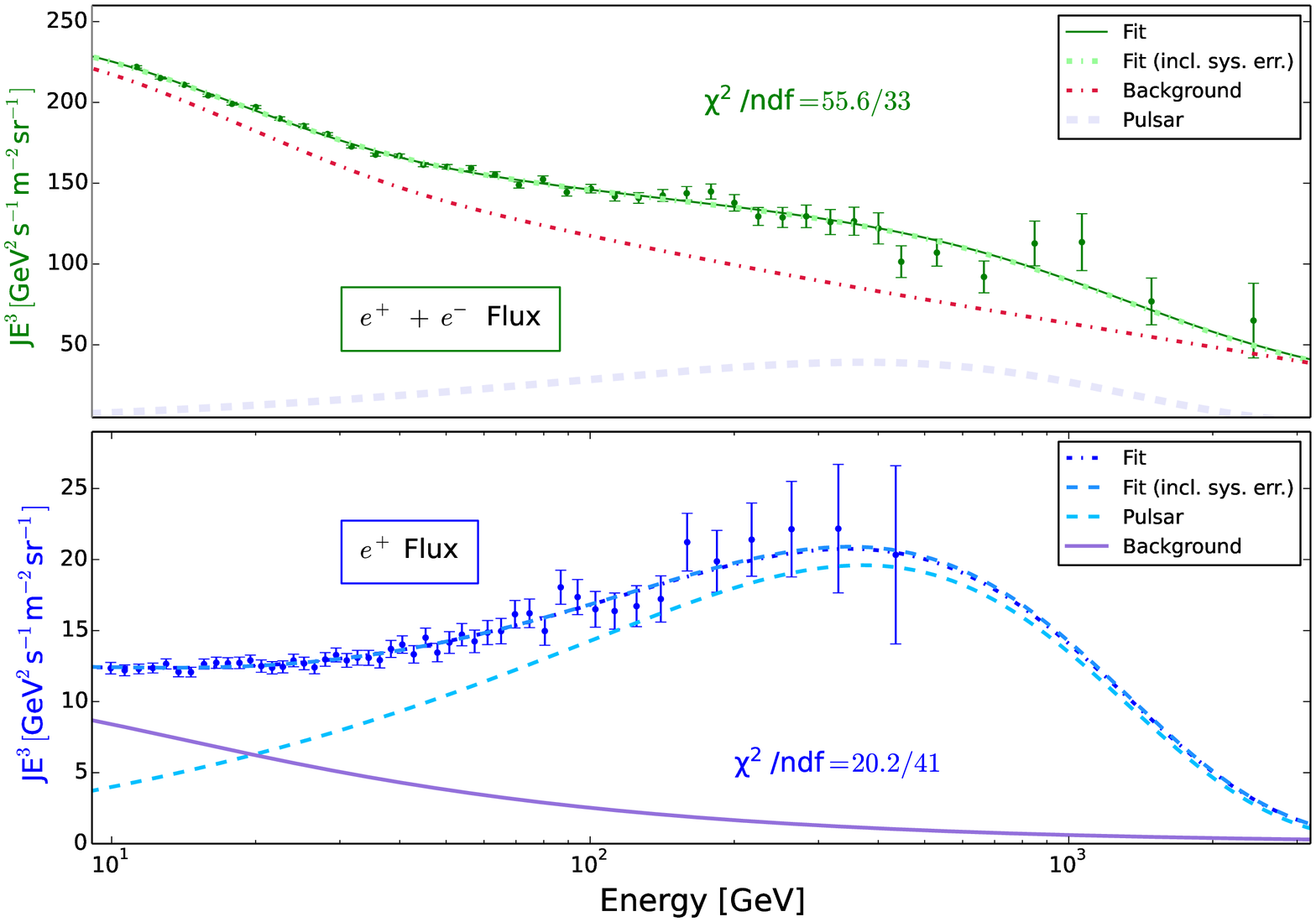} (a)\\%
\includegraphics[width=0.9\textwidth]{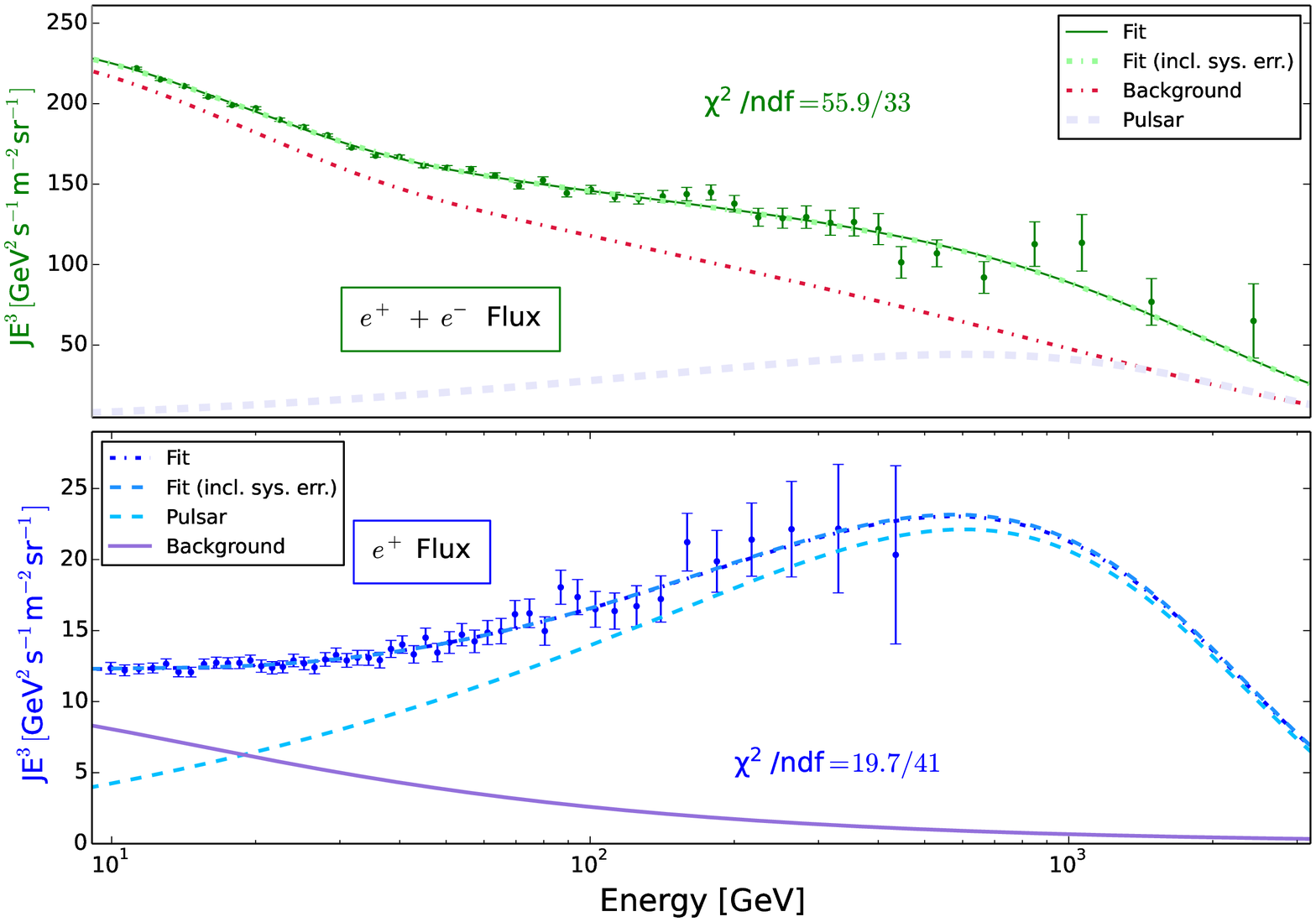} (b)
 \caption{Figure (a): Broken power law with pulsar model as extra source and $E_d$ set to 10~TeV, is taken as a test model (green line) to describe $e^+\, +\, e^-$ spectrum measured by CALET (upper panel) and $e^+$ flux measured by AMS-02 (black error bars) in the lower panel. The red-dotted line represents the background $e^+\, +\, e^-$ spectrum. Grey and cyan dashed lines in upper and lower panel show the contribution from the pulsar to the $e^+\, +\, e^-$ and $e^+$ spectrum, respectively. The best fit obtained including systematic errors is presented with green and blue dashed lines in the upper and lower panels, respectively. Figure(b): Same as figure (a), but with $E_d$ set to 2~TeV. The fit parameters are given in Table~\ref{Table:ModelCALETfitpulsar}.\label{fig:brkpowandsingpul}}
\end{figure}

With $E_d$ set to 10~TeV, the best fit pulsar model is obtained for $E_{\text{pwn}}=700$~GeV with $s=0.05$. With $E_d$ set to 2~TeV, the best fit values are $E_{\text{pwn}}=1.5$~TeV with $s=0.05$. A summary of the fit parameter values is given in Table~\ref{Table:ModelCALETfitpulsar} and the fitted curves are shown in figure~\ref{fig:brkpowandsingpul}. The secondary minima at $E_{\text{pwn}}\approx 200$~GeV in figure~\ref{fig:pulcutvschi}, for both $E_d=10$~TeV and $E_d=2$~TeV, are obtained for a smoothness value of 1. 


The power law index and cut-off energy of the pulsar parametrization in these best fit cases are comparable to the values in Ref.~\cite{Motz:2015cua} where this parametrization is compared with results of a numerical calculation of nearby pulsars. This indicates that the obtained values should correspond to the viable model of a young nearby pulsar like Monogem causing the positron excess. 

\begin{figure}
\includegraphics[width=0.85\linewidth]{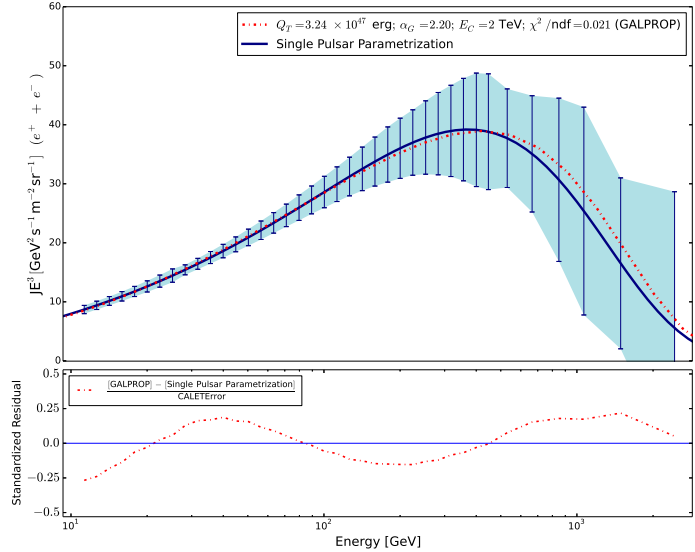}
\caption{Numerical propagation result from GALPROP for the Monogem pulsar is compared with the best case single pulsar model from the fit to CALET + AMS-02 for $E_d=10$~TeV. The blue error bars and shaded area represent the statistical error of the CALET measurement. In the lower panel we plot the standardized residual of the scaling fit and show that the difference is at most $\sim 0.25\sigma$. \label{fig:diffpul_Mon}}
\end{figure}

To evaluate the pulsar parameters associated with the fitted spectrum, we compare the power-law with cut-off extra source spectrum as obtained by the best-fit with the spectrum of the Monogem pulsar from numerical calculation with GALPROP. The $e^-\, (e^+)$~flux from the Monogem pulsar is calculated using the same propagation model as for the flux from DM decay, with the distance of Monogem from the solar system $d=0.28$~kpc and age $T=1.1\times 10^5$ years taken from the ATNF catalogue~\cite{Manchester:2004bp}. In the \mbox{GALPROP} calculation, the time progression is taken as $1.1\times 10^4$ steps of 10 years and the spatial grid separation is 0.1~kpc in a cube with a side length of 12~kpc centered on the solar system. The release of the CR is assumed to be instantaneous at the beginning of the pulsar's life, i.e. during the first 10 year time-step, and an exponentially cut off power-law spectrum is taken as the injection spectrum~\cite{Malyshev:2009tw}. The injection spectrum power-law index and cut-off energy were adapted by scanning to match the power-law with cut-off spectrum from the parametrization fit (eq.~\ref{eq:extrasourcepulsar}). The GALPROP results are also scaled to match the parametrization, with parameters of the best-fit model for $E_d=10$~TeV as shown in table~\ref{Table:ModelCALETfitpulsar}), and the best fitting case is shown in figure~\ref{fig:diffpul_Mon} with the parameters of - source injection index 2.20, source energy cut-off 2~TeV and total released energy  $3.24\, \times \, 10^{47}$~erg. These values are within the ranges commonly given in models for cosmic-ray acceleration in pulsars~\cite{Hooper:Pulsarposi, Hooper:DMpulsarplot, Malyshev:2009tw, Yuksel:2008rf, Feng:2015uta}. As shown in figure~\ref{fig:diffpul_Mon}, the difference between the result of GALPROP calculation and the parametrized spectrum is significantly smaller than the experimental errors, and thus the modeling of the propagated pulsar spectrum by the approximate power-law with cut-off parametrization is viable, given the precision and energy range of the studied experimental results.

The exponentially cut off pulsar spectrum is considerably different from the DM spectrum (figure~\ref{fig:800GeVDMEd2T}) which shows a harder drop in the flux at an energy corresponding to half the DM mass. In the next section we discuss the possibility of discerning the signatures of the pulsar spectrum from the DM decay, with five years of CALET measurement, based on simulated data.

\begin{table}
\tbl{Parameters obtained from the best fit of different studied models to combined CALET $e^+\, +\, e^-$ and AMS-02 $e^+$ spectra is listed here. Upper line (lower line) of each cell shows the values obtained without (with) including systematic error of CALET data. \label{Table:ModelCALETfitpulsar}}{
		\tabcolsep=0.05cm
		\begin{tabular}{|l|c|c|c|c|c|c|c|c|c|c|}\hline
		\makecell{\textbf{Model}\\ Pulsar}&\makecell{$C_{e^-}$\\$\left(\frac{\text{\tiny{GeV}}}{\text{s}\,\, \text{m}^{2}\,\, \text{sr}}\right)$}&$\frac{C_s}{C_{e^-}}$& $\gamma _{e^-}$& $\Delta\gamma _{e^-}$&\makecell{$E_g$\\ GeV}& $s$&\makecell{$E_{pwn}$\\\tiny{TeV}}&$\frac{C_{pwn}}{C_{e^-}}$&$\gamma _{pwn}$&$\chi^2$ \\	
		\multirow{6}{*}{$E_d=10~\text{TeV}$}&\multirow{4}{*}{$716$}&\multirow{4}{*}{$0.065$}&\multirow{4}{*}{$3.22$}&\multirow{4}{*}{$0.201$}&\multirow{4}{*}{$40.9$}&\multirow{4}{*}{$0.05$}&\multirow{4}{*}{0.7}&\multirow{4}{*}{0.0021}&\multirow{4}{*}{2.45}&\multirow{4}{*}{75.88}\\\cline{1-11}
		&&&&&&&&&&\\
		&\multirow{4}{*}{$719$}&\multirow{4}{*}{$0.065$}&\multirow{4}{*}{$3.23$}&\multirow{4}{*}{$0.203$}&\multirow{4}{*}{$40.7$}&\multirow{4}{*}{$0.05$}&\multirow{4}{*}{0.7}&\multirow{4}{*}{0.0021}&\multirow{4}{*}{2.47}&\multirow{4}{*}{$30.45$}\\\cline{2-11}
		&&&&&&&&&&\\
		&&&&&&&&&&\\\hline	
		\multirow{4}{*}{$E_d=2~\text{TeV}$}&\multirow{2}{*}{$705$}&\multirow{2}{*}{$0.059$}&\multirow{2}{*}{$3.20$}&\multirow{2}{*}{$0.224$}&\multirow{2}{*}{$40.3$}&\multirow{2}{*}{$0.05$}&\multirow{2}{*}{1.5}&\multirow{2}{*}{0.0027}&\multirow{2}{*}{2.53}&\multirow{2}{*}{$75.62$}\\
		&\multirow{4}{*}{$707$}&\multirow{4}{*}{$0.057$}&\multirow{4}{*}{$3.20$}&\multirow{4}{*}{$0.223$}&\multirow{4}{*}{$40.3$}&\multirow{4}{*}{$0.05$}&\multirow{4}{*}{1.5}&\multirow{4}{*}{0.0026}&\multirow{4}{*}{2.55}&\multirow{4}{*}{$30.86$}\\\cline{2-11}	
		&&&&&&&&&&\\
		&&&&&&&&&&\\\cline{1-11}
		\end{tabular}}
	\end{table}

\section{Discerning DM Decay and Pulsar Models with Expected CALET Data}\label{sec:5yrCALET}
Using a simulation of the expected data from five years of CALET observation, we now estimate the potential to distinguish the DM decay spectrum from the generic single pulsar spectrum. Based on the flux prediction from the fit of the parametrization with DM as extra source and assuming an aperture of $1040 \,\text{cm}^2\, \text{sr}$~\cite{PhysRevLett:CALET}, 5000 event samples were generated. These samples reflect the statistical fluctuation in the event rate, which causes various possible outcomes of the $e^+\, +\, e^-$ flux measurement in the DM decay scenarios. The pulsar source parametrization is then fitted to these CALET \mbox{$e^+\,+\, e^-$} samples and the AMS-02 $e^+$ flux measurement, obtaining a \mbox{$\chi ^2$ distribution}. In figure~\ref{fig:case507pulfittoDM}, we show an example of the single pulsar model fit to one of the 5000 CALET data samples, from the DM Model A as extra source. 
In this fitting, the background cut-off energy $E_d$ is chosen by lowest $\chi ^2$ from among the values 1~TeV, 2~TeV, 3~TeV, 5~TeV, 7~TeV and 10~TeV, to give the background spectrum the flexibility associated with the uncertainties on the propagation and source distribution. Similarly, the smoothness parameter $s$ is scanned from 0 to 1 in steps of 0.05, and the solar modulation potential $\phi$ from 0.35~GV to 0.65~GV in steps of 0.5~GV. 

In the same way, 5000 samples were created for each of the 800~GeV DM-decay models with low $\gamma$-ray yield, with example fits for the \mbox{DM Model B} and \mbox{DM Model C} also presented in figure ~\ref{fig:case507pulfittoDM}.
   
\begin{figure}
 \centering
 \includegraphics[width=0.98\linewidth]{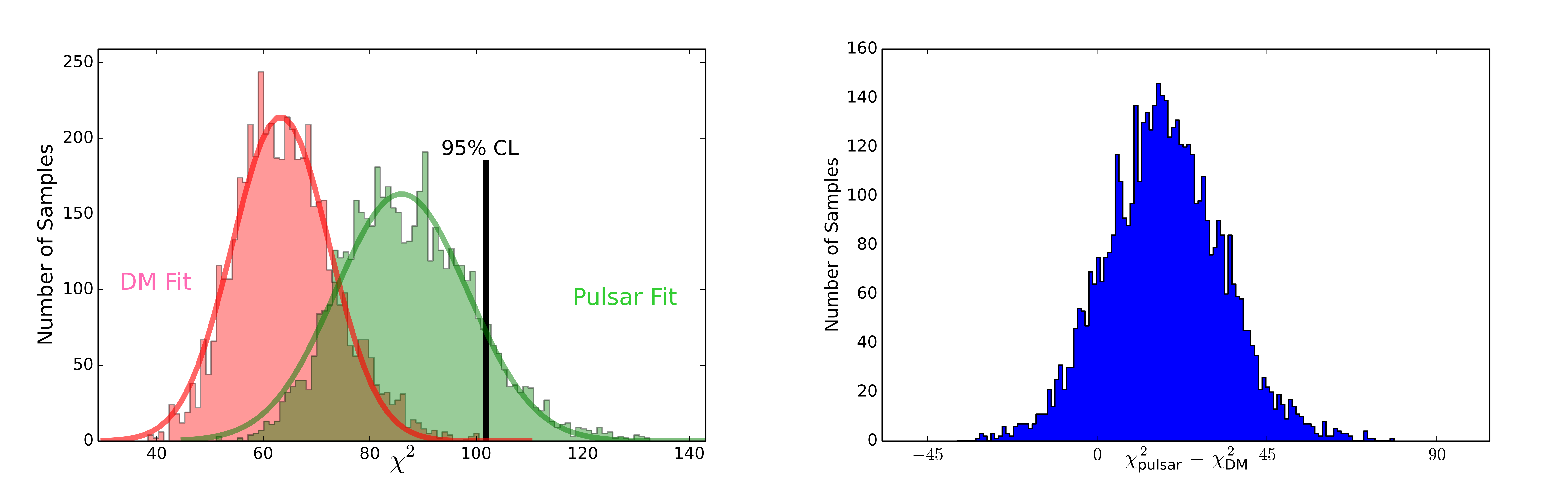} (a)\\%
 \includegraphics[width=0.98\linewidth]{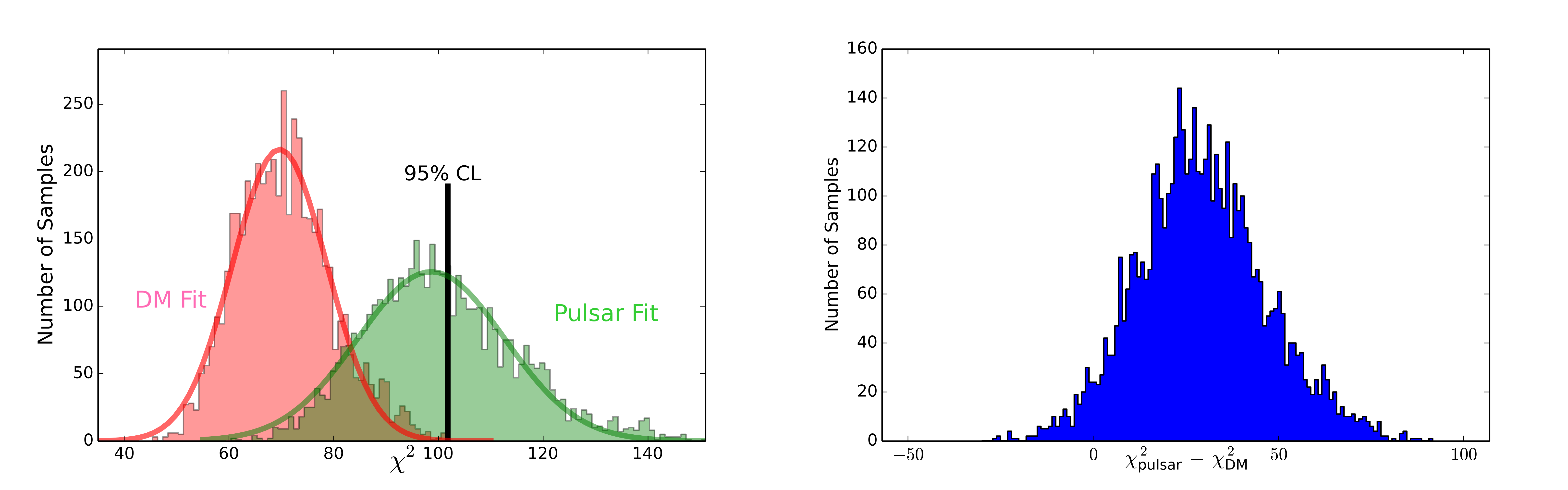} (b) \\%
 \includegraphics[width=0.98\linewidth]{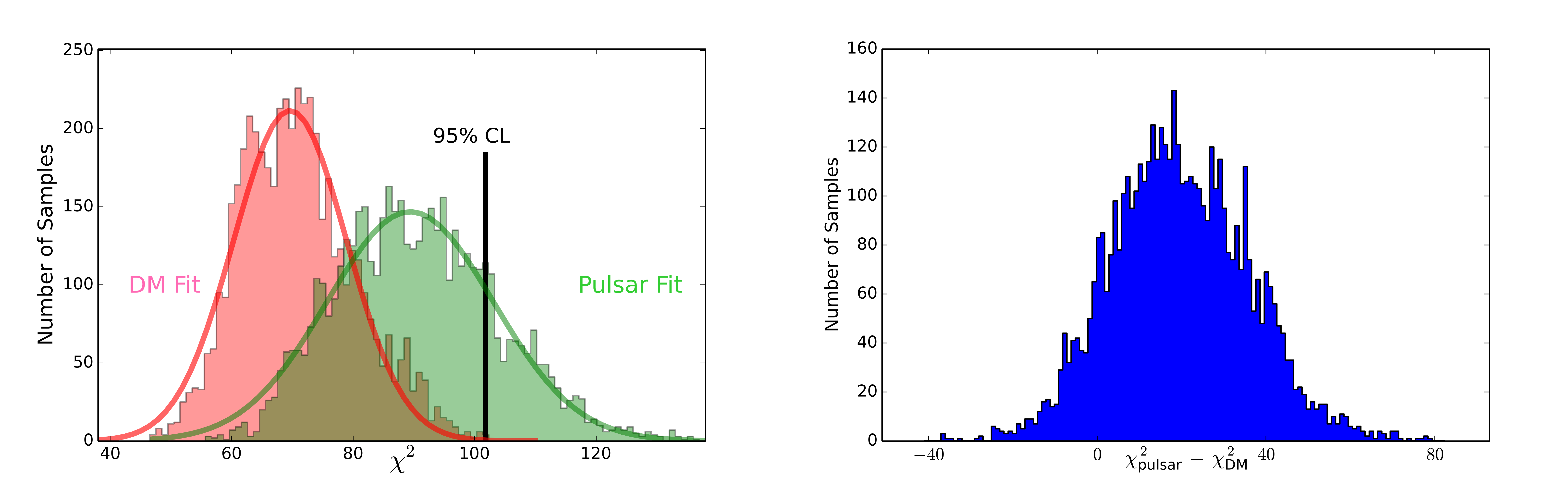} (c) %
 \caption{Figure (a): In left panel, $\chi ^2$ distribution for the fit of the single pulsar source to the simulated CALET data for 5000 DM Model A samples + AMS-02 positron flux data (green) and re-fit of DM case using the same samples (red). The $\chi ^2$ difference of pulsar fit and DM Model re-fit $\left(\chi ^2 _{\text{pulsar}} - \chi ^2 _{\text{DM}}\right)$ are shown in right panel. 
Similarly in figure (b) and figure(c), CALET data samples were generated assuming the DM Model B and DM Model C, with the $\chi ^2$ distribution and $\left(\chi ^2 _{\text{pulsar}} - \chi ^2 _{\text{DM}}\right)$ shown respectively.  \label{fig:0d8TDMbestcasehist}}
\end{figure}

The DM parametrization which was used to determine the models for which the CALET data samples were created is also fitted to the simulated data, yielding another $\chi ^2$ distribution for comparison. In figure~\ref{fig:0d8TDMbestcasehist}, we show the $\chi ^2$ distributions for the two scenarios, decaying DM and single pulsar source.
Since DM and pulsar model are non-nested i.e. the parameters describing these models are independent of each other, a quantitative likelihood ratio test is not conclusive. However the quality of the models to explain the measurement can be compared using Akaike's Information Criterion (AIC)~\cite{Akaikecri} as described in Ref.~\cite{Bhattacharyya:selfcite}. For a given set of models, the one with the lowest AIC value is most favorable to represent the data, if the likelihood for the models shows a normal distribution. The AIC value is defined as 
\begin{equation}
AIC = -2L_m + 2m \, ,    
\end{equation}
where $L_m$ is the maximum value of the log likelihood function and $m$ is the number of free parameters in the model. In addition to the free parameters of the background spectrum, both the DM and the pulsar model add three additional free parameters, reducing the comparison of the AIC to a comparison of $\chi ^2$. For the DM models, including the $\tau\tau\nu$ channel (which was ruled out in the fitting to build the models) the additional three free parameters are the decay times (eq.~\ref{eq:DMflux}) for the three lepton channels. For the pulsar model, the additional parameters are $C_{\text{pwn}},\, \gamma _{\text{pwn}},\, E_{\text{pwn}}$ from eq.~\ref{eq:extrasourcepulsar}. As can be seen from figure~\ref{fig:0d8TDMbestcasehist}, the $\chi ^2$ distribution for each  model follows a normal distribution. We plot the $\chi ^2$ difference between pulsar model fit and DM model re-fit $\left(\chi ^2 _{\text{pulsar}}\,-\,\chi ^2 _{\text{DM}}\right)$  on the right side of figure~\ref{fig:0d8TDMbestcasehist}, and for all 3 DM models, it is positive for $\sim 90\%$ of the samples. According to the AIC, for most samples the initially simulated DM model is thus favored to represent the CALET \mbox{$e^+\, +\, e^-$} simulated flux and \mbox{AMS-02} $e^+$~flux over the hypothetical pulsar model. 

In particular, if the pulsar model is excluded at $95\,\%$ CL, while the DM model is below this threshold, it can be concluded that the DM model can eventually be distinguished by the CALET measurement from the pulsar case. The number of excluded samples for fitting the pulsar model, and for re-fitting of the DM model, are listed in table~\ref{Table:CALETfitDMandpul} for the three simulated DM models. As the lower tail of the $\chi ^2$ distribution of the pulsar fit may be slightly shortened because of the discretization of $\phi,\, E_d$ and $s$, a Gaussian curve is fitted to the histograms to obtain a more precise and conservative value for the percentage of excluded samples.
It is shown that the re-fit of the DM model has a good fit-quality in each case. An equally good $\chi ^2$ distribution is obtained, if excluding the $\tau\tau\nu$ channel in the re-fitting of the DM model, though the model has one less free parameter, proving that the combination of simulated CALET data and AMS-02 data represents the DM model well.  

It is found that less than half of the pulsar samples would be excluded at $\chi ^2 > 95\%$~CL for the three studied DM models based on the simulated future CALET data. 
However, $\left(\chi ^2 _{\text{pulsar}}\,-\,\chi ^2 _{\text{DM}}\right)$ is $\sim 20$ or more on average for a clear majority of the samples, indicating that better separation could be possible with an analysis using more advanced statistical methods for non-nested models, for example following the principle used in Ref.~\cite{Algeri:2015zpa}, and/or focusing on the region of interest near the cut-off energy of the extra source. 

\begin{table}
\tbl{Based on 5000 CALET data samples simulated for each DM model, the number of samples that are excluded at $95\%$~CL from the fit of pulsar case and DM case to the simulated CALET $e^+\, +\, e^-$ data + AMS-02 $e^+$ data are listed, as well as the fraction of excluded samples according to the fit of a normal distribution to each histogram. Samples with positive $\chi ^2$ difference between pulsar model and DM Model re-fit are listed for the studied DM Models. \label{Table:CALETfitDMandpul}}{
		\tabcolsep=0.05cm
		\begin{tabular}{|c|c|c|c|c|c|c|c|c|}\hline
		\multirow{2}{*}{\makecell{\textbf{Model}\\ DM}}& \multicolumn{3}{|c|}{\makecell{Pulsar Case Excluded Sample}}& \multicolumn{3}{|c|}{\makecell{DM Case Wrongly Excluded Sample}}&\multicolumn{2}{|c|}{\makecell{$\chi ^2 _{\text{pwn}} - \chi ^2 _{\text{DM}}>0$}} \\\cline{2-9}	 
		&\makecell{Sample\\ Number}&Fraction&\makecell{Fraction\\(Gaussian Fit)}&\makecell{Sample \\ Number}&Fraction&\makecell{Fraction\\(Gaussian Fit)}&\makecell{Sample\\Number}& \makecell{Average\\ value}\\\cline{1-9}
		\multirow{2}{*}{\makecell{Model A}}&&&&&&&&\\
		&\multirow{1}{*}{$621$}&$12.42\%$&\multirow{1}{*}{$12.03\%$}&\multirow{1}{*}{$0$}&$0\%$&\multirow{1}{*}{$1\times 10^{-3}\%$}&\multirow{1}{*}{4188}&$17.82$\\
		&&&&&&&&\\\hline
		\multirow{2}{*}{\makecell{Model B}}&\multirow{2}{*}{$2237$}&\multirow{2}{*}{$44.74\%$}&\multirow{2}{*}{$44.21\%$}&\multirow{2}{*}{$0$}&\multirow{2}{*}{$0\%$}&\multirow{2}{*}{$1\times 10^{-2}\%$}&\multirow{2}{*}{$4791$}&\multirow{2}{*}{$29.35$}\\
		&&&&&&&&\\
		&&&&&&&&\\\hline
		\multirow{2}{*}{\makecell{Model C}}&\multirow{2}{*}{$1077$}&\multirow{2}{*}{$21.54\%$}&\multirow{2}{*}{$20.78\%$}&\multirow{2}{*}{$0$}&\multirow{2}{*}{$0\%$}&\multirow{2}{*}{$2\times 10^{-2}\%$}&\multirow{2}{*}{$4459$}&\multirow{2}{*}{$19.53$}\\
		&&&&&&&&\\
		&&&&&&&&\\\hline		
		\end{tabular}}
\end{table}

\begin{landscape}\centering
 \begin{figure}\centering
 \begin{minipage}[t]{.3\textwidth}
 \centering
 \vspace{2.7cm}
 \includegraphics[width=66mm, height=84mm]{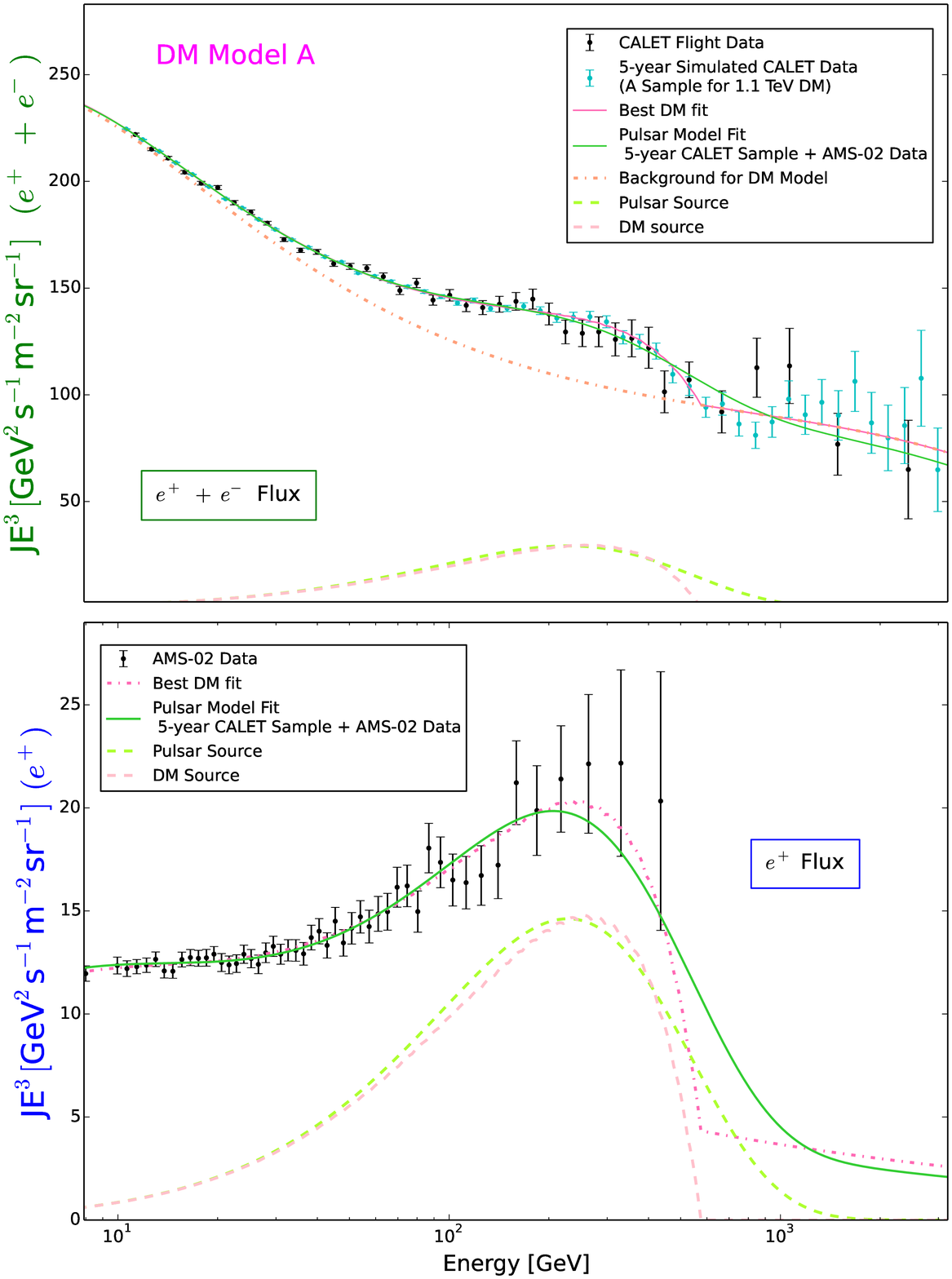} (a)
 \end{minipage}
 \hfill
 \begin{minipage}[t]{.3\textwidth}
 \centering
 \vspace{2.7cm}
 \includegraphics[width=66mm, height=84mm]{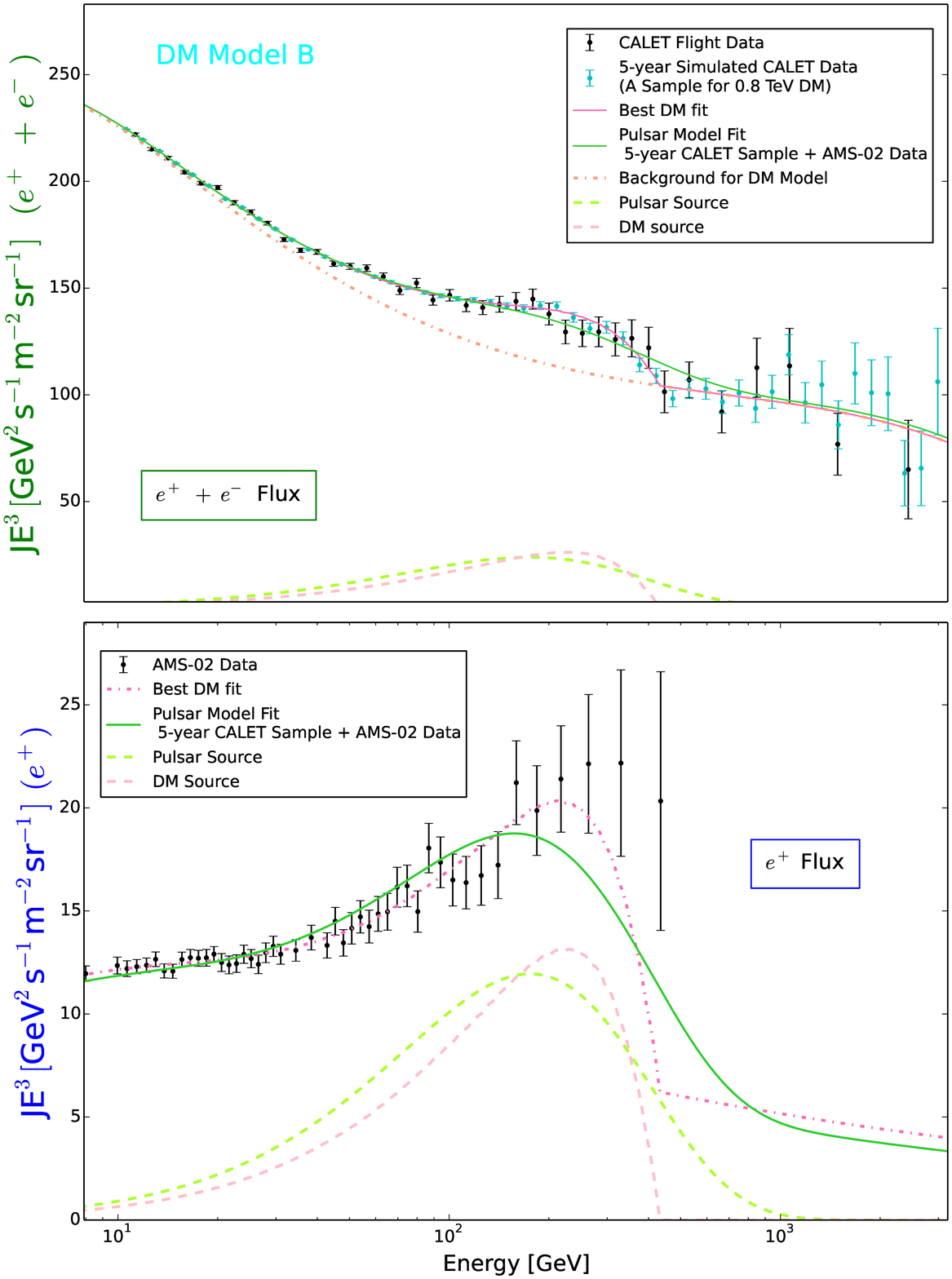} (b)
 \end{minipage}
 \hfill
 \begin{minipage}[t]{.3\textwidth}
 \centering
 \vspace{2.7cm}
 \includegraphics[width=64mm, height=84mm]{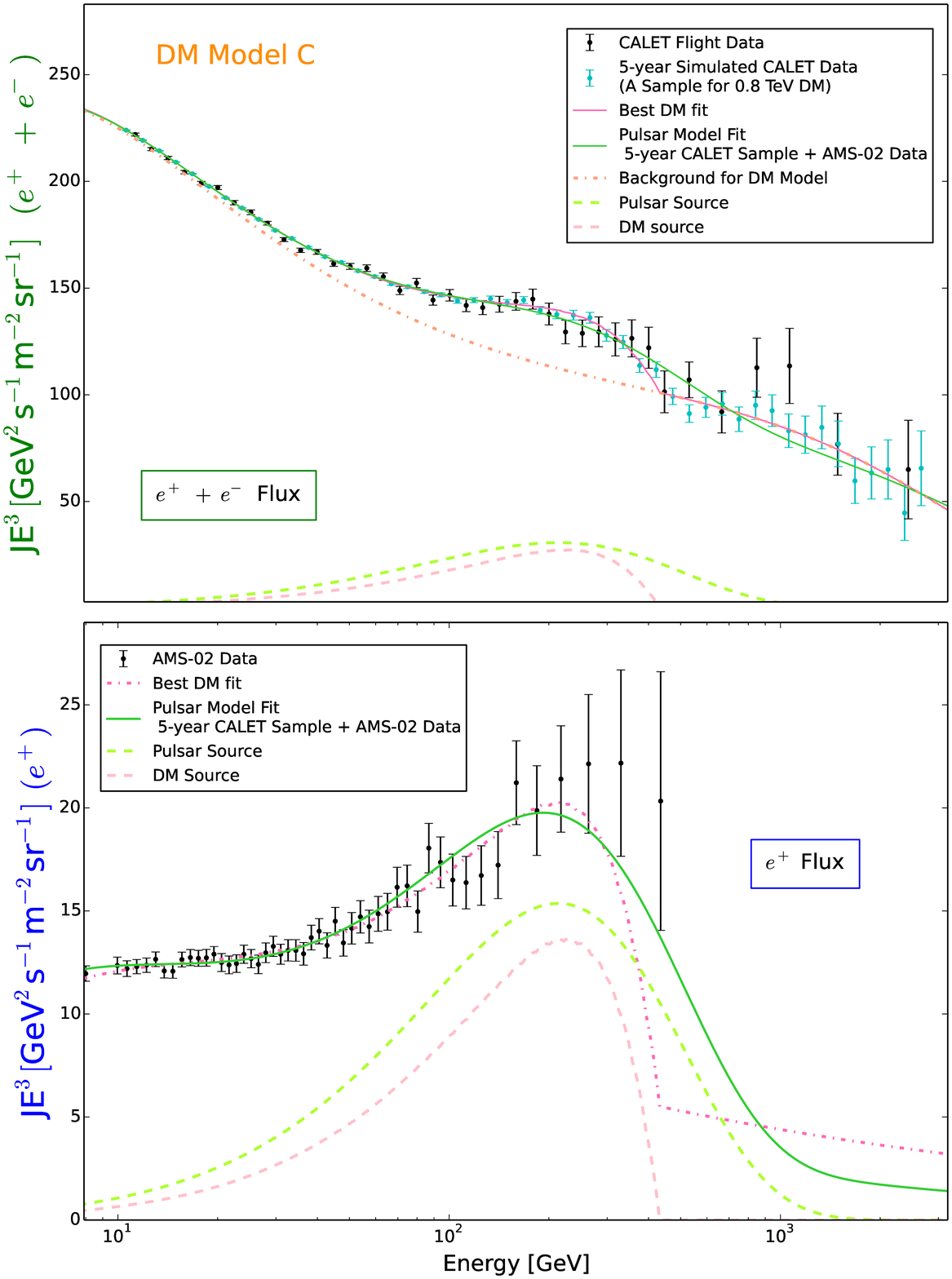} (c)
 \caption{Figure (a): Fit of the single pulsar source (green line) to one of the 5000 statistical samples (cyan dots) of five year CALET measurement for $e^+\, +\,e^-$ flux (upper panel) assuming DM Model A as extra source and $e^+$ flux measured by AMS-02 (lower panel) data is shown here. Dotted lines in the upper panel represents the background spectrum for the DM source. Pulsar source and DM decay contribution in the $e^+\, +\,e^-$ flux and $e^+$ flux are shown with green and pink dashed lines respectively. Black error-bars represent CALET and AMS-02 flight data. Similar plots where DM Model B and DM Model C are used to simulate CALET data are shown in figure (b) and figure (c) respectively.\label{fig:case507pulfittoDM}}
 \end{minipage}
 \vfill
 \end{figure}
\end{landscape}

\section{Conclusion}

First results of the $e^+\, +\, e^-$ spectrum measured by CALET up to 3~TeV were published recently. 
In this work we performed a combined analysis of this CALET result and the $e^+$~flux measured by AMS-02. With a parametrization of the background assuming a smoothly broken power-law for the primary electron flux, and either 3-body decay of DM or a pulsar as the extra source causing the positron excess, we show that both pulsar and DM model can well explain these measurements. The ranges of DM mass and the exponential cut-off energy of the pulsar model which, under the chosen conditions, could explain the combined CALET and AMS-02 measurements are shown. 
The $\gamma$-ray emission caused by the DM decay in these models is not exceeding the Fermi-LAT diffuse $\gamma$-ray flux. Finally, from the analysis of simulated data for five years of CALET observation, it is shown that a separation of the DM models from the pulsar scenario is possible with a maximum probability of $\sim 45\%$ for DM Model B and less for DM Model A and \mbox{DM Model C}, based on the used simple statistical method. 
A refined analysis based on the $\chi ^2$ difference, which is found to be $\sim 20$ on average, is promising to achieve a better probability to discern the two scenarios.


\bibliographystyle{ws-ijmpd}
\bibliography{reference}

\end{document}